\documentclass{aa}  

\usepackage{graphicx}
\usepackage{txfonts}
\usepackage{siunitx,subfig,xspace,natbib}
\usepackage{hyperref}
\usepackage{amsmath}%
\usepackage{amsfonts}%
\usepackage{amssymb}%
\usepackage{mathrsfs}
\usepackage{arydshln}
\usepackage{cprotect}

\DeclareMathOperator\erfc{erfc}

\newcommand{\mrm}[1]{\mathrm{#1}}

\begin{document}

   \title{Constraints on positron annihilation kinematics in the inner Galaxy}


\author{
  Thomas Siegert   		     \inst{\ref{inst:mpe},\ref{inst:xcu},\ref{inst:ucsd}}\thanks{E-mail: tsiegert@mpe.mpg.de} \and
	Roland M. Crocker        \inst{\ref{inst:anu}} \and
  Roland Diehl             \inst{\ref{inst:mpe},\ref{inst:xcu}} \and
	Martin G. H. Krause      \inst{\ref{inst:hfs}} \and
	Fiona H. Panther         \inst{\ref{inst:anu},\ref{inst:unsw},\ref{inst:arc}} \and
	Moritz M. M. Pleintinger \inst{\ref{inst:mpe}} \and
	Christoph Weinberger     \inst{\ref{inst:mpe}}
}
\institute{
  Max-Planck-Institut f\"ur extraterrestrische Physik, Gie\ss enbachstra\ss e, D-85741 Garching, Germany
  \label{inst:mpe}
  \and
  Excellence Cluster Universe, Boltzmannstra\ss e 2, D-85748, Garching, Germany
  \label{inst:xcu}
	\and
	Center for Astrophysics and Space Sciences, University of California, San Diego, 9500 Gilman Dr, La Jolla, CA 92093, USA
	\label{inst:ucsd}
	\and
	Research School of Astronomy and Astrophysics, Australian National University, Canberra 2611, A.C.T., Australia
  \label{inst:anu}
	\and
	Centre for Astrophysics Research, School of Physics, Astronomy and Mathematics, University of Hertfordshire, College Lane, Hatfield, Hertfordshire AL10 9AB, UK
  \label{inst:hfs}
	\and
	School of Science, UNSW Canberra, Australian Defence Force Academy, Canberra 2612, Australia
	\label{inst:unsw}
	\and
	ARC Centre of Excellence for All-Sky Astrophysics (CAASTRO)
	\label{inst:arc}
}

   \date{Received July 12, 2018; accepted May 31, 2019}

 
  \abstract
   {The annihilation of cosmic positrons with electrons in the interstellar medium results in the strongest persistent $\gamma$-ray line signal in the sky. For the past 50 years, this 511~keV emission - predominantly from the galactic bulge region and from a low surface-brightness disk - has puzzled observers and theoreticians.  A key issue for understanding positron astrophysics is found in cosmic-ray propagation, especially at low kinetic energies ($\lesssim 10$~MeV).}
   {We want to shed light on how positrons propagate and the resulting morphology of the annihilation emission. We approach this "positron puzzle" by inferring kinematic information of the 511~keV line in the inner radian of the Galaxy. This constrains propagation scenarios and positron source populations in the Milky Way.}
   {By dissecting the positron annihilation emission as measured with INTEGRAL/SPI, we derive spectra for individual and independent regions in the sky. The centroid energies of these spectra around the 511~keV line are converted into Doppler-shifts, representing the line-of-sight velocity along different galactic longitudes. This results in a longitude-velocity diagram of positron annihilation. From high-resolution spectra, we also determine Doppler-broadening from $\gamma$-ray line shape parameters to study annihilation conditions as they vary with galactic longitude.}
   {We find line-of-sight velocities in the 511~keV line that are consistent with zero, as well as with galactic rotation from CO measurements ($2$--$3~\mrm{km~s^{-1}~deg^{-1}}$), and measurements of radioactive $\mrm{^{26}Al}$ ($7.5$--$9.5~\mrm{km~s^{-1}~deg^{-1}}$). The velocity gradient in the inner $\pm30^{\circ}$ is determined to be $4\pm6~\mrm{km~s^{-1}~deg^{-1}}$. The width of the 511~keV line is constant as a function of longitude at $2.43\pm0.14$~keV, with possibly different values towards the disk. The positronium fraction is found to be 1.0 along the galactic plane.}
   {The weak signals in the disk leave open the question whether positron annihilation is associated with the high velocities seen in $\mrm{^{26}Al}$ or rather with ordinarily rotating components of the Milky Way's interstellar medium. We confirm previous results that positrons are slowed down to the 10~eV energy scale before annihilation, and constrain bulk Doppler-broadening contributions to $\lesssim 1.25$~keV in the inner radian. Consequently, the true annihilation conditions remain unclear.}


   \maketitle
%

\section{Introduction}

The interpretation of the morphology of $\gamma$-ray emission from positron annihilation in the Milky Way has remained in contention since the discovery of the galactic 511~keV line in the late 1960's \citep{Haymes1969_511,Johnson1972_511}. Unlike at any other wavelength, the bulge region dominates the signal, with a flux ratio between bulge and disk of 0.6 \citep{Siegert2016_511} \citep[see also][]{Milne1997_511,Knoedlseder2005_511,Bouchet2010_positron}. Depending on the distance assumed for bulge and disk, this converts to a luminosity ratio between 0.3 and 1.0. While many plausible astrophysical sources of positrons are concentrated in the thin disk of the Galaxy, such as from massive stars and their core-collapse supernovae, the 511~keV map is more reminiscent of an old stellar population, such as type Ia supernovae. With increased observing time from the $\gamma$-ray spectrometer SPI \citep{Vedrenne2003_SPI} aboard the INTEGRAL satellite \citep{Winkler2003_INTEGRAL}, it has become clearer that the 511~keV disk of the Milky Way is probably not truncated beyond $20^{\circ}$ longitude, as found using data from OSSE aboard CGRO \citep{Purcell1997_511}, but rather extended in both longitude \citep{Bouchet2010_511,Skinner2014_511} and latitude \citep{Siegert2016_511}. In general, the connection between the positrons seen to annihilate and their source regions within the Galaxy, and the number of positrons actually produced in the possible sources, are the main questions to be understood.

The transformation from supposedly disk-dominated production sites to a bulge-dominated annihilation region inevitably suggests propagation of positrons over larger distances, once they managed to escape their production sites. Alternatively, the positron sources could be distributed like the annihilation radiation, which might be surprising, but is not ruled out. For all of the candidate sources and the production mechanisms, positrons start out at relativistic kinetic energies: From $\beta^+$-decays of nucleosynthesis ejecta, the kinetic energies are restricted to $\lesssim 10~\mrm{MeV}$, with important astrophysical $\beta^+$-decayers, such as $\mrm{^{26}Al}$ or $\mrm{^{44}Ti}$, providing typically $\lesssim 1~\mrm{MeV}$ \citep{Endt1990_nuclei21-44}. If electrons and positrons are created in high energy-density environments, such as around black holes or neutron stars, even higher kinetic energies, of the order of GeV, can be achieved. Observational constraints from direct annihilation in-flight of relativistic positrons with electrons at rest \citep{Aharonian1981_TPA,Svensson1982_ann_spec} suggest injection energies below $3$~MeV \citep{Beacom2006_511}. Similar to other cosmic-ray species, positrons can also be boosted in energy from diffuse shock acceleration and obtain very high energies, if they do not annihilate inside the shock region \citep{Ellison1990_positrons}.

\begin{figure}[!ht]%
\centering
\includegraphics[width=0.9\columnwidth,trim=3.0cm 1.7cm 1.8cm 2.4cm,clip=true]{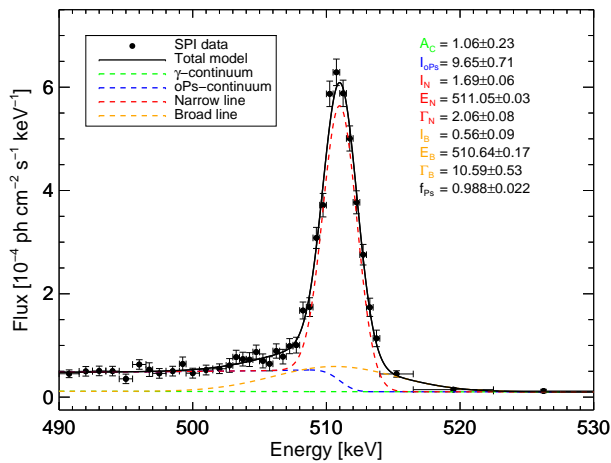}%
\caption{Positron annihilation spectrum of the Galaxy using the best-fit morphology model of \citet{Siegert2016_511}. Shown are the SPI data points in black ($\geq 3\sigma$-bins), and four model components: narrow 511~keV line (red), broad 511~keV line (orange), ortho-positronium continuum (blue), galatic $\gamma$-ray continuum (green). The total model is shown as solid black line. The fitted parameters are shown in the legend. The units are $10^{-5}~\mrm{ph~cm^{-2}~s^{-1}~keV^{-1}}$ for the continuum amplitude $A_C$, $10^{-3}~\mrm{ph~cm^{-2}~s^{-1}}$ for the line fluxes $I_N$ and $I_B$, and the positronium continuum $I_{oPs}$, and keV for the lines centroids $E_N$ and $E_B$, and widths $\Gamma_N$ and $\Gamma_B$. The positronium fraction $f_{Ps}$ is measured to be between $0.97$ and $1.00$.}%
\label{fig:gal511spec}%
\end{figure}

The propagation of charged particles in the interstellar medium (ISM) is guided by the galactic magnetic field and bulk plasma flows, and governed by energy losses or gains due to specific conditions of temperature, density, field strengths, or ionisation state. The $\gamma$-ray observations made with INTEGRAL/SPI are consistent with positrons annihilating predominantly in warm ($T \approx 7000$--$40000$~K, or $0.6$--$3.5$~eV) and partly ionised ($x \approx 0.01$-$0.20$) gas phases \citep[e.g.][]{Jean2006_511,Churazov2005_511,Churazov2011_511,Siegert2017_PhD}. This could be the ISM or stellar atmospheres \citep{Murphy2005_posiloss,Bisnovatyi-Kogan2017_511}. Note, however, that positron annihilation on dust is also a possible interpretation \citep{Guessoum2005_511,Guessoum2010_511}. Positrons mainly form positronium (Ps) before annihilating (Ps-decay), which is only efficient below a kinetic energy of $\approx 1$~keV. At higher energies, the cross-section for positron interactions in the ISM is dominated by excitation and ionisation interactions with atoms. Consequently, positrons slow down from relativistic energies, experiencing both continuous energy losses and binary interactions with atoms in the ISM until they reach nearly zero (i.e. thermal) energies. Positron annihilation in the Galaxy occurs on a quasi-steady time scale. Once they are produced at a specific rate, depending on the source variety, they annihilate within a characteristic time, depending on the large- and small-scale structure of the Milky Way, and the prevailing conditions. This makes observing annihilation $\gamma$-rays valuable for probing the physics of low-energy cosmic-rays.

These findings ignored the impact of the large-scale dynamics of the Milky Way which results in a Doppler-broadening of the Galaxy-wide 511~keV line. This adds to the line broadening due to the true annihilation conditions in a particular region. Depending on the true kinematics, the intrinsic 511~keV line width from the the ISM parameters alone may be smaller:

If the 511~keV kinematics closely follow those of galactic rotation at about $\approx 220~\mrm{km~s^{-1}}$ as measured from CO emission, this would point to the classical scenario in which positron initially start at MeV energies, propagate through the ISM, thereby cool and slow down, and annihilate in a favourable phase. Here, the energy loss must be rapidly increased with respect to the environment of the positron sources, as otherwise, the positronium fraction would be smaller and the rotation velocity would hardly be kept. Already thermalised positrons, on the other hand, can only undergo charge exchange with neutral H in the far end of the Maxwell-distribution, also resulting in 511~keV Doppler-shifts as would be expected from the bulk motion of neutral or molecular gas in the Galaxy. Such a scenario would result in a Doppler-broadening of the line up to 0.75~keV.

About 5--10\% of the Milky Way's positron budget (5--15\% in the disk) can be explained \citep{Knoedlseder1997_PhD,Siegert2017_PhD} by decaying $\mrm{^{26}Al}$ from massive stars. The kinematics from the $\mrm{^{26}Al}$ 1808.63~keV $\gamma$-ray line are peculiar in that they show a significant, $200~\mrm{km~s^{-1}}$, excess in the line-of-sight velocities with respect to CO \citep{Kretschmer2013_26Al}. The authors suggest that massive star ejecta stream preferably into regions with a smaller density (less pressure), away from the molecular clouds where the massive stars have formed, and towards galactic rotation. This makes $\mrm{^{26}Al}$ appear to rotate faster. Even though the decay-positrons from $\mrm{^{26}Al}$ share their velocities ($\beta^+$-decay is angle-independent), they have to slow down again first, which would probably be in the surrounding HI shells, and again at galactic CO rotation speeds. A 511~keV velocity profile that shows similar absolute values would only be possible if special conditions are met. Then, additional broadening of up to 1.2~keV may occur.

A flat rotation curve in the inner Galaxy would point to a completely different annihilation region, kinematically distinct from the disk. Dispersion of the annihilation sites may dominate over rotation in the 511~keV bulge, maybe suggesting a link to the galactic halo. RR Lyrae stars, for example, which are commonly found in globular clusters \citep[e.g.][]{Clement2001_RRL,Clementini2001_RRL} and in the galactic centre \citep{Baade1946_RRLgalcen} show a flat velocity profile in the inner $\pm 5^{\circ}$ of Milky Way, suggesting them to originate from an inner-halo sample \citep{Kunder2016_lvstars}. If positrons annihilate in or close to their sources \citep[e.g. in clumpy/dusty ejecta; ][]{Zhang2014_RCorBor_WDmerger,Jeffers2012_RCorBor_dust}, another galactic-wide positron production and annihilation scenario connected to the old stellar population may be considered. \citet{Crocker2017_511_91bg} discussed the merger of a pure He with a CO white dwarf to be the progenitor channel of peculiar SN1991bg-like supernovae \citep{Pakmor2013_SNIa_merger,Dan2015_SNIa_merger} in which major amounts of $\beta^+$-unstable $\mrm{^{44}Ti}$ are produced \citep{Perets2010_SN2005E_44Ti,Waldman2011_SN2005E_44Ti}. Alternatively, a signal without kinematics with respect to the solar system could also be interpreted as an annihilation site nearby, with a source feeding it steadily. The Doppler-broadening of any such scenario would be small, $\lesssim 0.25$~keV.

Additional trends in spectral appearance of the 511~keV line along galactic longitudes could furthermore reveal distinct kinematic structures and separate annihilation regions. Consistent annihilation flux, line width, and line centroid variations may point to known large-scale components of the Milky Way, such as the galactic bar \citep[e.g.][]{Wegg2013_bulgebar,Ellis2018_leptonium}, or rather local features such as the local bubble \citep[e.g.][]{Lallement2014_OB}. Note that even if positrons are already thermalised and avoided forming Ps, they undergo elastic scattering \citep{Gryzinski1965_particle_collisions}, which can potentially broaden the 511~keV line in addition.

By investigating the spectral characteristics, especially the kinematics, of the 511~keV line in different regions of the Galaxy, we want to shed light on the link between positron sources, their propagation, and their final annihilation. We exploit the spatial and spectral resolution of SPI to measure the kinematics as a function of longitude. With such a study, the conditions of annihilation during propagation can be approached from a different perspective and could thus provide independent insights about this long-lasting "positron puzzle". Our SPI measurements combine many lines-of-sight, and the resulting velocity measurements reflect the radial velocity of the Milky Way, weighted by the three-dimensional 511~keV photon density distribution. 

This paper is structured as follows: In Sect.~\ref{sec:data_analysis}, the general data analysis methods for the SPI coded-mask telescope are presented (Sect.~\ref{sec:data_general}), and how kinematic information from positron annihilation spectra are extracted (Sects.~\ref{sec:simultan_fit} and \ref{sec:lbv-diagram}). The results, i.e. the spectral parameters as a function of longitude, are presented in Sect.~\ref{sec:results}. We discuss the 511~keV kinematics and their implications to annihilation conditions in Sect.~\ref{sec:discussion}, and provide a summary in Sect.~\ref{sec:summary}.

\section{Data analysis}\label{sec:data_analysis}

\subsection{Data and general method}\label{sec:data_general}

We use the INTEGRAL/SPI data set from \citet{Siegert2016_511}, in which the large-scale diffuse annihilation emission was investigated. This comprises 73590 pointings, targeting predominantly the bulge and the disk of the Milky Way, for a total exposure of 189~Ms. This exposure is reduced to 160~Ms due to failures and processing time of the detectors ("dead time"). In total, the number of spectra in the data set is $1,214,799$, taking into account dead detectors.

In general, SPI data are analysed by comparing measured counts $d_k$ per energy bin $k$ over the time of the exposure, i.e. time sequences, with predicted time sequences of celestial emission plus background models. These components add up to the total model $m_k$, with

\begin{equation}
m_k = \sum_j R_{jk} \sum_{i=1}^{N_S} \theta_{i} M_{ij} + \sum_{t_B} \sum_{i = N_S+1}^{N_S + N_B} \theta_{i,t_B} B_{ik}\mrm{.}
\label{eq:modeldesc}
\end{equation}

Here $N_S$ is the number of sky models, $M_{ij}$ for each image element $j$, to which the imaging response function (coded mask pattern) $R_{jk}$ is applied. The $N_B$ background models $B_{ik}$ are added to the celestial emission, and can in general be time-dependent. The expected detector patterns for sky and background are known \citep[see e.g.][]{Siegert2016_511,Siegert2017_PhD}, and only the amplitudes $\theta_{i,(t_B)}$ are determined in a maximum likelihood fit. The background model has also been adopted from \citet{Siegert2016_511}. As photon counting obeys the Poisson statistics, we make use of the Cash statistics,

\begin{equation}
C(D|\theta_i) = 2 \sum_k \left[ m_k - d_k \ln m_k \right]\mrm{,}
\label{eq:cstat}
\end{equation}

i.e. one version of the log-likelihood of the Poisson statistics \citep{Cash1979_cstat}, to optimise the fit and to find the intensity scaling parameters $\theta_{i,(t_B)}$. In particular, we consider complementary sky models, described in Sect.~\ref{sec:multi-fit} (see also Appendix \ref{sec:slidingwindow}), one or more defining the regions for which the spectrum is to be extracted, and one describing the remaining 511~keV sky. \citet{Siegert2016_511} used four smooth Gaussian-shaped components to model the galactic-wide annihilation emission. Because this model may not represent the true morphology, but may vary on smaller angular scales, the intensity scaling parameters per energy bin, $\theta_{i,(t_B)}$, are re-determined for each sky region and the background.

\begin{figure}[!ht]%
\centering
\includegraphics[width=0.9\columnwidth]{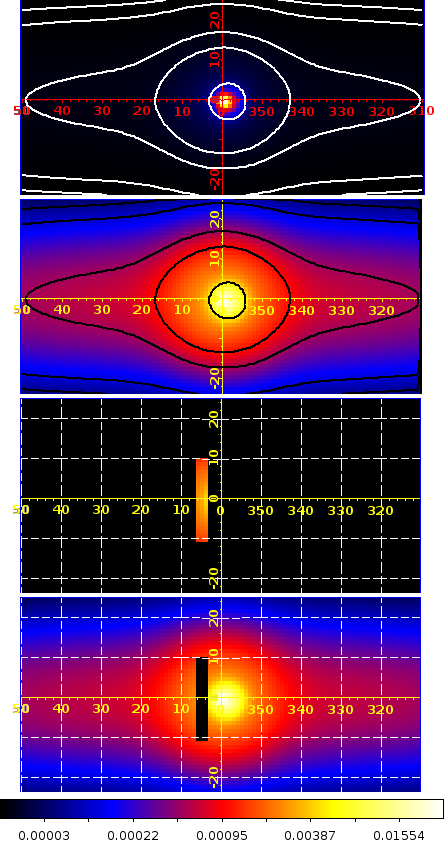}%
\caption{Celestial emission model components for galactic 511~keV radiation. From top to bottom: Galactic 511~keV line map as determined in \citet{Siegert2016_511} in linear scaling; same as shown in the top panel, but in logarithmic scaling to emphasise low surface-brightness regions; exemplary ROI region of $\Delta l \times \Delta b = 3.00^{\circ} \times 21.00^{\circ}$ size, centred at $l=4.75^{\circ}$; complementary map, excluding the ROI region. The contours indicate regions in which 20\%, 50\%, 80\%, 98\%, and 99.7\%, respectively, of the total flux is contained (from inside to outside).}%
\label{fig:sliding_window}%
\end{figure}

The celestial emission in this energy range is dominated by the bright 511~keV line emission from the Galaxy's centre, modelled by a narrow bulge, a broad bulge, and a point-like source centred at $(l/b) = (0^{\circ}/0^{\circ})$. To this, a Gaussian-shaped disk is added to capture the faint and low surface-brightness emission beyond the bulge. The details of the spatial emission models are found in \citet{Siegert2016_511}, and in Appendix~\ref{sec:appendix_morphology}.

The instrumental background is modelled in a self-consistent way, using long-term knowledge about the detector behaviours, space-craft conditions, and physical origins of $\gamma$-ray emissions in the satellite \citep{Siegert2016_511}. Based on the INTEGRAL/SPI BackGround and Reponse Data Base \citep[BGRDB][]{Diehl2018_SPI}, the background detector patterns (relative counts of the detectors to each other) are predicted as a function of energy and time. This data base includes spectral information of hundreds of instrumental $\gamma$-ray lines which are associated to different isotopes in the satellite, being excited by cosmic rays and de-exciting either promptly or delayed. The continuum background, mainly due to bremsstrahlung inside the spacecraft, is modelled in the same manner. Long-term trends of detector degradation from this cosmic-ray bombardment and radioactive build-ups are individually traced by this technique. In this energy range, eight instrumental $\gamma$-ray lines, on top of a power-law shaped instrumental continuum is used to model these background patterns \citep{Siegert2017_PhD}, to be distinguished from the expected sky emission patterns. The background scaling parameters are determined on a three-day interval \citep[see also][]{Siegert2019_SPI}, corresponding to one INTEGRAL orbit. This amounts to 1083 background scaling parameters for each, the continuum and line backgrounds as a whole, for a total of 2166 background parameters.

In summary, 6 (2) parameters are fitted for the celestial emission in a simultaneous (consecutive) fit of the regions of interest and the remaining Galaxy, and 2166 parameters for the instrumental background. The number of degrees of freedom in each individual energy bin thus amounts to $1,212,627$ ($1,212,631$).

\subsection{Spectral extraction}\label{sec:simultan_fit}

\subsubsection{Full sky 511~keV signal}\label{sec:full_sky}

The total spectrum of the 511~keV template map (Fig.~\ref{fig:sliding_window}, top two panels) is shown in Fig.~\ref{fig:gal511spec}, together with a multi-component fit. We identify four spectral components: a narrow and a broad 511~keV line, an ortho-positronium (o-Ps) continuum, and the galactic diffuse $\gamma$-ray continuum. The derived spectral parameters are consistent with previous studies, although both, the broad and the narrow annihilation line, appear to be broader than previously measured \citep{Jean2006_511}. This may arise from different line shapes in the disk and the bulge \citep{Siegert2016_511}. The measured line width of $2.06\pm0.08~\mrm{keV}$ (FWHM above instrumental resolution, narrow line) is supposedly a superposition of Doppler-shifts from galactic rotation and intrinsic broadening from the annihilation conditions, i.e. how and where the positrons annihilate. Doppler-shifts alone would imply a rotation velocity of the order of $1500~\mathrm{km~s^{-1}}$. Since galactic rotation may contribute to this line broadening, the 511~keV line width from annihilation conditions only could appear narrower. Hence in this study, we separate the kinematic from the intrinsic line broadening due to annihilation conditions. A discussion of annihilation conditions should always consider the kinematic broadening as well.

\subsubsection{Segmenting the 511~keV sky}\label{sec:multi-fit}

In order to extract a position-velocity diagram from INTEGRAL/SPI data, we dissect the full-sky 511~keV emission model into six parts: Five components define our regions of interest (ROIs), each inside a spherical rectangle of $l \in [l_0^r - \Delta l / 2,l_0^r + \Delta l / 2]$ and $b \in [b_0 - \Delta b / 2,b_0 + \Delta b / 2]$, and its complement, consisting of the remaining 511~keV map with those regions cut out. Here, $l_0^r \in \{ -23.75^{\circ}, -11.75^{\circ}, +0.25^{\circ}, +12.25^{\circ}, +24.25^{\circ} \}$ is the centre longitude of each ROI, with a longitudinal width of $\Delta l = 12^{\circ}$ for a high spectral resolution analysis, and $b_0 = -0.25^{\circ}$ is the latitude centre, identical for all ROIs, with a latitudinal extent of $\Delta b = 21^{\circ}$ (ROI set \texttt{L1}). At a distance of $R_0 = 8.5~\mrm{kpc}$ to the galactic centre, the ROI width of $\Delta b = 21^{\circ}$ converts to a characteristic scale height $h_0$, given by $h_0 = |R_0 \tan(\Delta b)|$. A large scale height of $\approx 1$~kpc was suggested by the study of \citet{Siegert2016_511}. For an estimate of how large the impact of the choice of the longitude bin centres is, we perform analyses with four different ROI sets in broad longitude bins, and three different ROI sets in narrow longitude bins. This provides an estimate of the systematic uncertainties of the measured parameters, and at the same time might reveal trends which may be hidden by an inappropriate binning (Appendix~\ref{sec:systematics}). The longitudinal width of the ROIs is chosen to account for the angular resolution (point spread function (PSF) width of $2.7^{\circ}$ FWHM) and the field of view ($16^{\circ}\times16^{\circ}$) of SPI, coupling different emission regions beyond the PSF width in its tails \citep[cf.][]{Attie2003_SPI}. This large field of view thus requires a full description of the remaining 511~keV sky, as typical observations of diffuse emission combine photons from all over this solid angle covered by SPI. Any ROI $r$ is hence expressed as

\begin{eqnarray}
M_{ROI}^r(l,b) & = & \frac{M(l,b)}{A_{tot}} \times A_{ROI}^r \times \nonumber \\
	& \times & \left[ \Theta(l -  l_0^r + \Delta l / 2) - \Theta(l - l_0^r - \Delta l / 2) \right] \times \nonumber \\
	& \times& \left[ \Theta(b -  b_0 + \Delta b / 2) - \Theta(b -  b_0 - \Delta b / 2) \right]\mrm{.}
\label{eq:ROI_math_definition}
\end{eqnarray}

A single small ROI is illustrated in Fig.~\ref{fig:sliding_window}, bottom two panels. The remaining complementary map is then

\begin{equation}
	M_{COMP}(l,b) = A_{COMP} \times \left(M_{511}(l,b) - \sum_{r=1}^{5} M_{ROI}^r(l,b)\right)\mrm{.}	
\label{eq:complementary_map_full}
\end{equation}

In Eqs.~(\ref{eq:ROI_math_definition}) to (\ref{eq:complementary_map_full}), $\Theta(x)$ is the Heaviside-function, and $A_{ROI}^r$ and $A_{COMP}$ are the fitted parameters in the maximum likelihood fit, Eq.~(\ref{eq:modeldesc}). In this way, the sky is represented in full spatial detail, and the intensity scaling parameters are fitted to SPI data for the entire set of ROIs and the entire complementary map.

We use the best-fitting multi-component sky model as derived by \citet{Siegert2016_511} as a galactic 511~keV emission model, see Eq.~(\ref{eq:gauss_2d2}). Here, we choose a celestial region $-50^{\circ} < l < +50^{\circ}$, $-25^{\circ} < b < +25^{\circ}$, to perform our analysis, as more than 90\% of the emission is included in this area. The Crab and Cygnus X-1 are located outside this region. We provide detailed investigations of all spectral parameters as a function of longitude in Sect.~\ref{sec:results}, and test the consistency of our results also as a function of latitude (see Appendix~\ref{sec:systematics}).

\begin{figure}[!ht]%
\centering
\includegraphics[width=0.9\columnwidth,trim=0.86in 1.06in 1.49in 0.49in,clip=true]{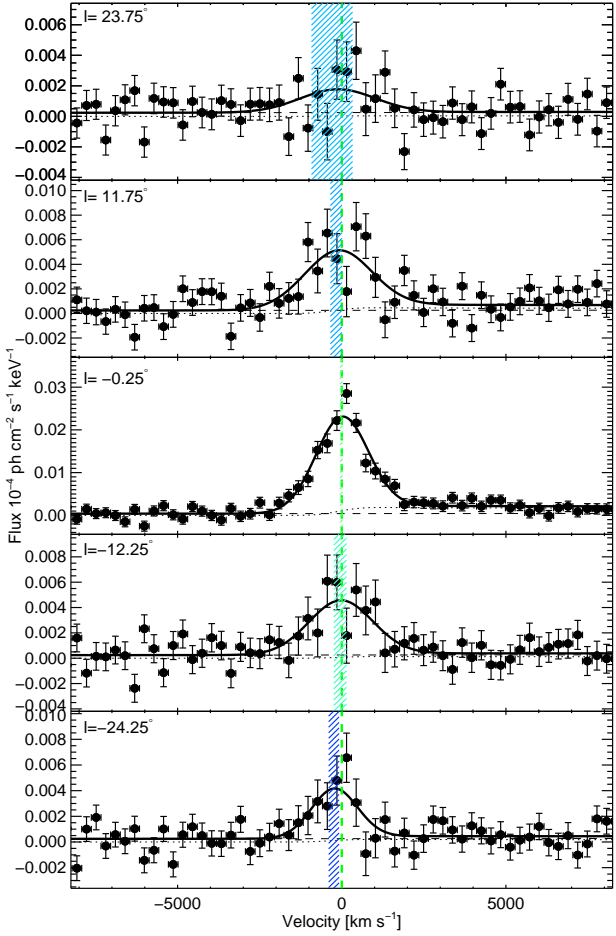}%
\cprotect\caption{Compilation of extracted ROI spectra (black data points, $1\sigma$ errors bars) in velocity space for different longitude regions (annotation in the upper left). Each spectrum has been fitted with the spectral model of Eq.~(\ref{eq:fit_fun_total}, thick solid line). Different spectral components are the diffuse continuum (dashed line), and the ortho-positronium continuum (dotted line), and the degraded Gaussian line with the fitted Doppler-shifts and their uncertainties marked by the hatched areas. These indicate either blue-shift, red-shift, or no shift. A trend from positive to negative longitudes, as could be expected from galactic rotation, is absent. Note that the y-axis scale varies between panels.}%
\label{fig:specs_L1}%
\end{figure}

This method ignores prior information on the data, such as the absolute normalisation of the individual ROIs, which have been determined already in a previous study \citep{Siegert2016_511}. Simultaneously splitting the emission map into many ROIs increases the statistical uncertainties in the derived spectra, because the number of parameters fitted is larger than actually required. As an alternative to this approach, we describe the "sliding window" method to analyse SPI Doppler-shift data in Appendix~\ref{sec:slidingwindow}, as it has been used by \citet{Kretschmer2013_26Al} to determine the longitude-velocity diagram of decaying $\mrm{^{26}Al}$ at 1808.63~keV. We compare the different analysis variants in Appendix~\ref{sec:method_comparison}.

\begin{figure}[!ht]%
\centering
\includegraphics[width=0.9\columnwidth,trim=0.75in 0.68in 0.94in 0.96in,clip=true]{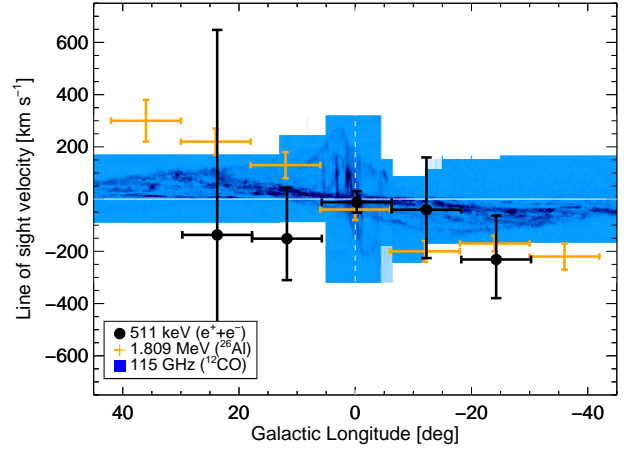}%
\cprotect\caption{Longitude-velocity diagram of the 511~keV positron annihilation line in the central radian of the Milky Way. In black, independent data points and one-sigma measurement uncertainties are shown. In addition, the Doppler-velocity measurements of decaying $\mrm{^{26}Al}$ at 1808.63~keV from \citet{Kretschmer2013_26Al} (orange), and line of sight velocity from CO measurements \citep{Dame2001_galvel} (blue-shaded areas) are shown for comparison.}%
\label{fig:lvdiagram}%
\end{figure}

\subsection{Position-velocity diagram}\label{sec:lbv-diagram}

For the five ROIs, the spectra (cf. Fig.~\ref{fig:specs_L1}) represent the maximum likelihood fits energy bin by energy bin, according to Eqs.~(\ref{eq:modeldesc}) and (\ref{eq:cstat}). We characterise these spectra individually through the 511~keV line position and width via a fit with a degraded Gaussian function, $L(E)$, on top a power-law-shaped continuum $C(E)$, and the o-Ps continuum $O(E)$. In particular, the spectra are modelled via

\begin{equation}
F(E) = C(E) + I_{L} \left( \frac{L(E)}{I_L} + R_{OL} \frac{O(E)}{I_O} \right)
\label{eq:fit_fun_total}
\end{equation}

which is similar to the fitting function of \citet{Jean2006_511}. In Eq.~(\ref{eq:fit_fun_total}), $I_{L}$ and $I_{O}$ are the total integrated fluxes of the 511~keV line and the o-Ps continuum, respectively, and $R_{OL} \equiv I_{O}/I_{L}$ is the ratio between the two. The functions have the following functional form:

\begin{equation}
C(E;C_0,\alpha) = C_0 \left(\frac{E}{511~\mrm{keV}}\right)^{\alpha}
\label{eq:conti}
\end{equation}

\begin{align}
G(E;A_0,E_0,\sigma) & = A_{0} \exp \left(- \frac{(E-E_{0})^2}{2\sigma} \right) \label{eq:gaussian} \\
T(E;\tau) & = \frac{1}{\tau} \exp \left( - \frac{E}{\tau} \right) \quad \forall E > 0 \label{eq:tail} \\
L(E;A_0,E_0,\sigma,\tau) & = (G \otimes T)(E) \label{eq:cls_function}
\end{align}

\begin{equation}
O(E;O_0,E_0) = 2 O_0 \left( T_1 - T_2 \ln(T_3) + T_4 + T_5 \ln(T_3) \right)
\label{eq:ortho_function}
\end{equation}

In Eq.~(\ref{eq:conti}), $C_{0}$ is the continuum amplitude, normalised to 511~keV, and $\alpha$ is the power-law index. Eq.~(\ref{eq:cls_function}) describes a degraded Gaussian function with a low-energy tail to account for detector worsening due to cosmic-ray bombardment. It is derived from a physical model of a perfectly symmetric Gaussian response, $G(E)$, with amplitude $A_0$, width $\sigma$, and centroid $E_0$, convolved with an exponential tail function, $T(E)$, with degradation parameter $\tau$, to describe the loss in the collection efficiency of the charge carriers. The o-Ps spectrum, Eq.~(\ref{eq:ortho_function}), with amplitude $O_0$, was first derived by \citet{Ore1949_511}, and is also convolved with the spectral response function of SPI. The fitted parameters in each spectrum are $C_{0}$, $O_0$, $A_0$, $\sigma$, and $E_0$. The $\gamma$-ray continuum power-law index is fixed to $-1.7$ \citep[cf.][]{Kinzer2001_511,Jean2006_511}, and the degradation parameter $\tau$ is fixed to the mean value over the data set, $\tau = 0.15$~keV. The full analytical expressions can be found in Appendix~\ref{sec:appendix_functions}, together with additional derived parameters of the spectral shape, such as the integrated line flux, $I_L$, and width (FWHM, $\Gamma_L$), the Ps flux $I_O$, and the Ps fraction, $f_{Ps}$. The latter describes the fractions of positrons undergoing a bound state - Ps - before annihilation (Ps decay). 

In each fit, $\sigma$ is constrained to a minimum width of $0.91$~keV (instrumental resolution of $2.15$~keV at 511~keV). We use the Metropolis-Hastings algorithm \citep{Metropolis1953_MH_algorithm,Hastings1970_MH_algorithm} to perform maximum likelihood fits of each spectrum. We use the parameter distributions for each parameter to calculate the expectation value as the best fit value, and the 68.3\% interval around this value as the fit parameter uncertainties. Examples of how the fits perform are shown by corner plots in Appendix~\ref{sec:corner_plots}. We obtain adequate individual fits to the spectra, based on reduced $\chi^2$ goodness-of-fit measures between 0.8 and 1.3 \citep[75 degrees of freedom;][]{Andrae2010_chi2}. The Doppler-velocities are estimated by determining the peak positions of the line function, Eq.~(\ref{eq:cls_function}). This requires the evaluation of the first derivative of the function,

\begin{equation}
\left. \frac{\partial L(E)}{\partial E}\right|_{E=E_{peak}} \stackrel{!}{=} 0\mrm{,}
\label{eq:epeak_full}
\end{equation}

which reduces to

\begin{equation}
E_{peak} \approx E_0 - \tau
\label{eq:approxpeak}
\end{equation}

for small values of $\tau$. For a line width of instrumental resolution, i.e. for $(\sigma/\tau) = (0.91/0.15)~\mrm{keV}$, the approximation is accurate to better than $1~\mrm{km~s^{-1}}$. The line peak positions are accurate to $\pm1$~keV or less ($<0.2\%$), so we can use the non-relativistic Doppler-formula to estimate the line-of-sight velocity,

\begin{equation}
v_{los} = \frac{E_{lab} - E_{peak}}{E_{lab}} c\mrm{.}
\label{eq:Dopplerformula}
\end{equation}

Here, $E_{lab} = 510.999~\mrm{keV}$ is the rest-mass energy of the electron, and $c = 299792.458~\mrm{km~s^{-1}}$ is the speed of light. This definition leads to negative line-of-sight velocities of blue-shifted lines, and positive velocities for red-shifted lines. By consecutively converting the measured Doppler-shifts in longitude-ROIs to line-of-sight velocities, we create a longitude-velocity (l-v) diagram of positron annihilation in the inner Galaxy. In Fig.~\ref{fig:lvdiagram}, the data points for the 511~keV line are shown, and compared to other Doppler-shift data (see Sects.~\ref{sec:rotation}). We estimate the total systematic uncertainties to be of the order of $100~\mrm{km~s^{-1}}$ ($|l| \lesssim 5^{\circ}$) to $200~\mrm{km~s^{-1}}$ ($|l| \gtrsim 5^{\circ}$). This is mainly driven by the 511~keV morphology in the inner disk, and the flux variations in individual ROIs; see Appendix~\ref{sec:systematics} for further details. 

\begin{figure*}[!ht]
	\centering
		  \subfloat[Integrated 511~keV line flux $I_L$. \label{fig:l_flux_plot}]{\includegraphics[width=0.89\columnwidth,trim=0.74in 0.70in 0.99in 0.97in,clip=true]{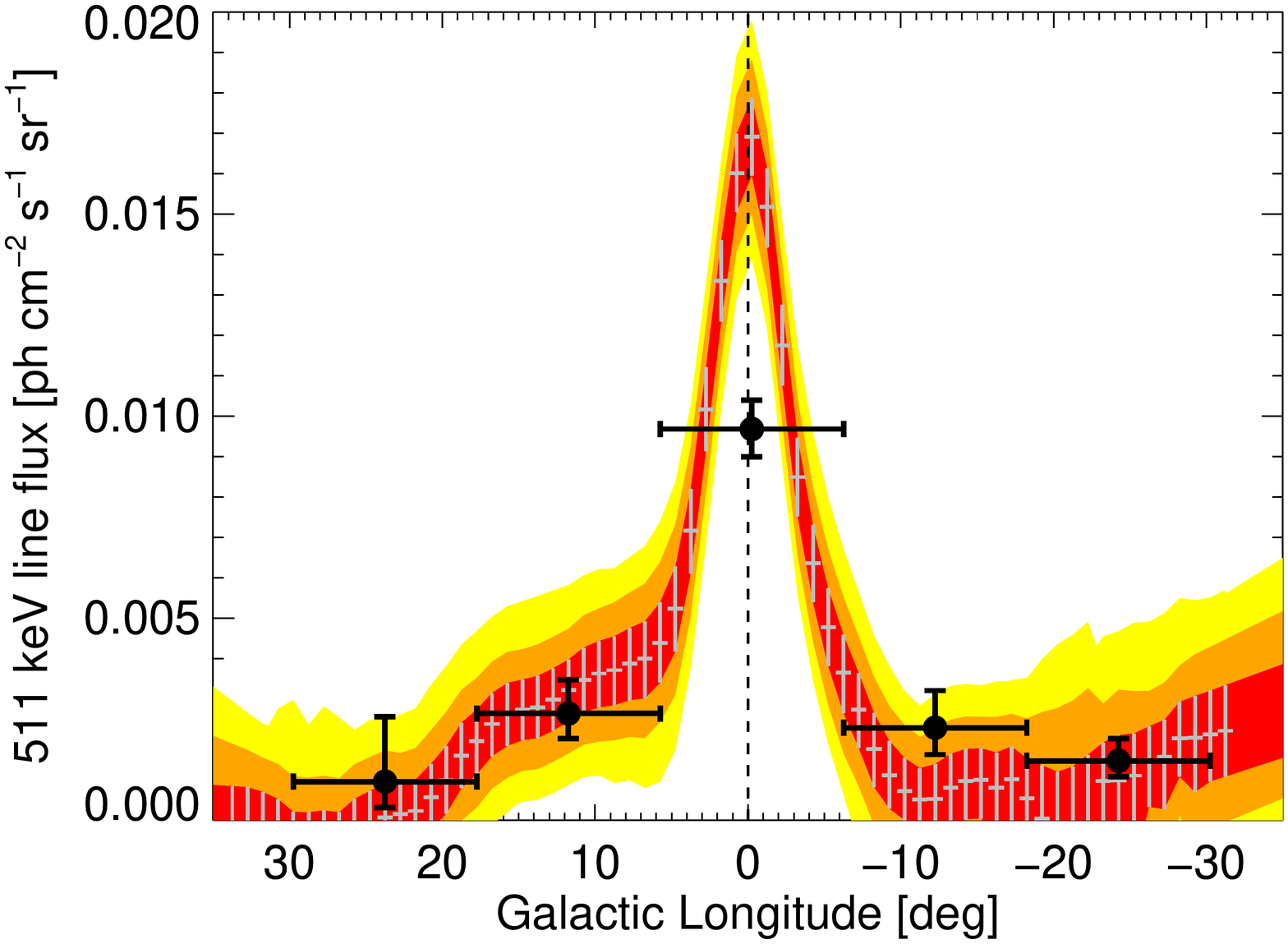}}~
			\subfloat[Line of sight Doppler-velocity $v_{los}$. \label{fig:l_v_plot}]{\includegraphics[width=0.89\columnwidth,trim=0.74in 0.70in 0.99in 0.97in,clip=true]{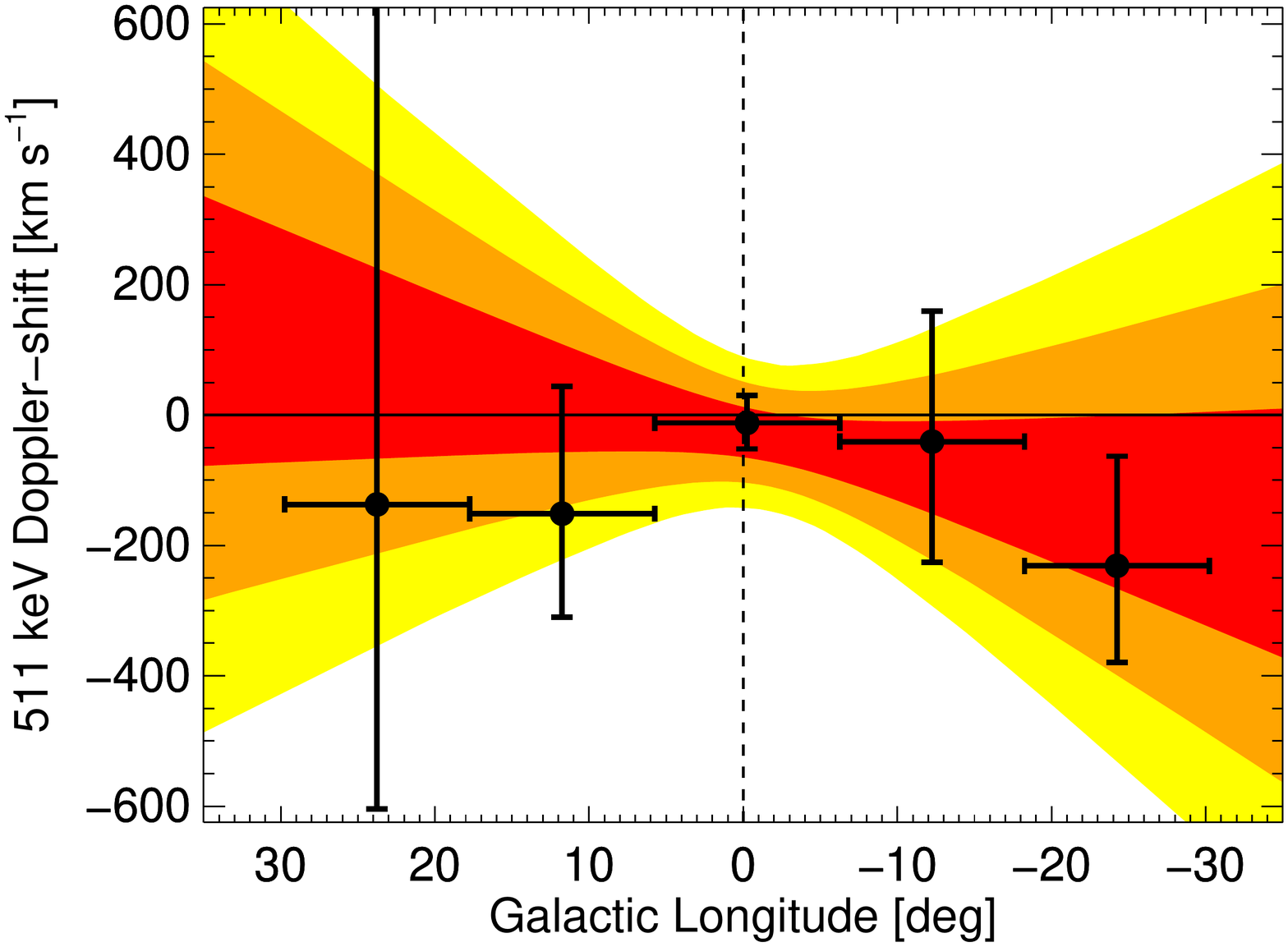}}\\
		  \subfloat[Astrophysical FWHM $\Gamma_L$. \label{fig:l_fwhm_plot}]{\includegraphics[width=0.89\columnwidth,trim=0.74in 0.70in 0.99in 0.97in,clip=true]{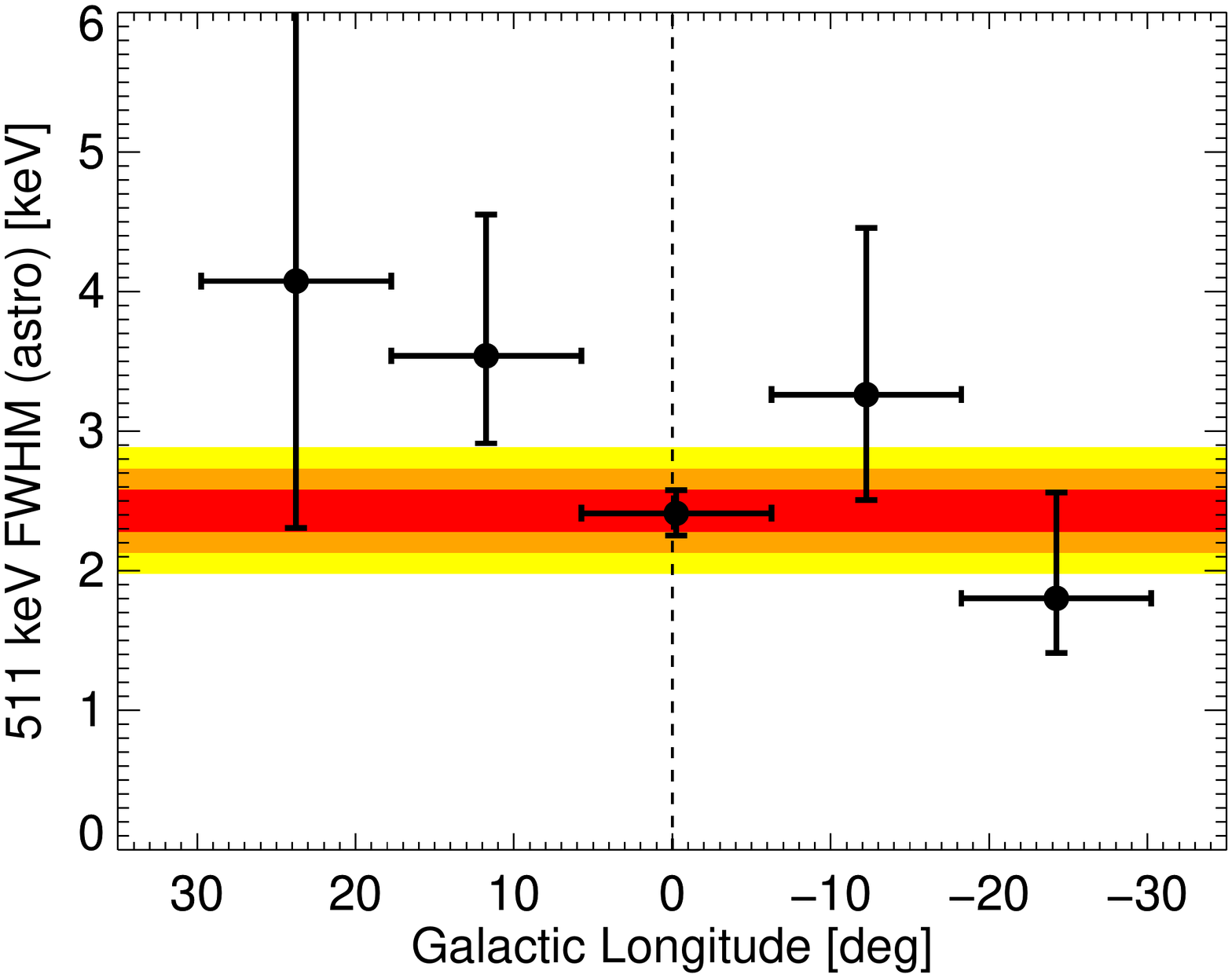}}~
			\subfloat[Positronium fraction $f_{Ps}$. \label{fig:l_fps_plot}]{\includegraphics[width=0.89\columnwidth,trim=0.74in 0.70in 0.99in 0.97in,clip=true]{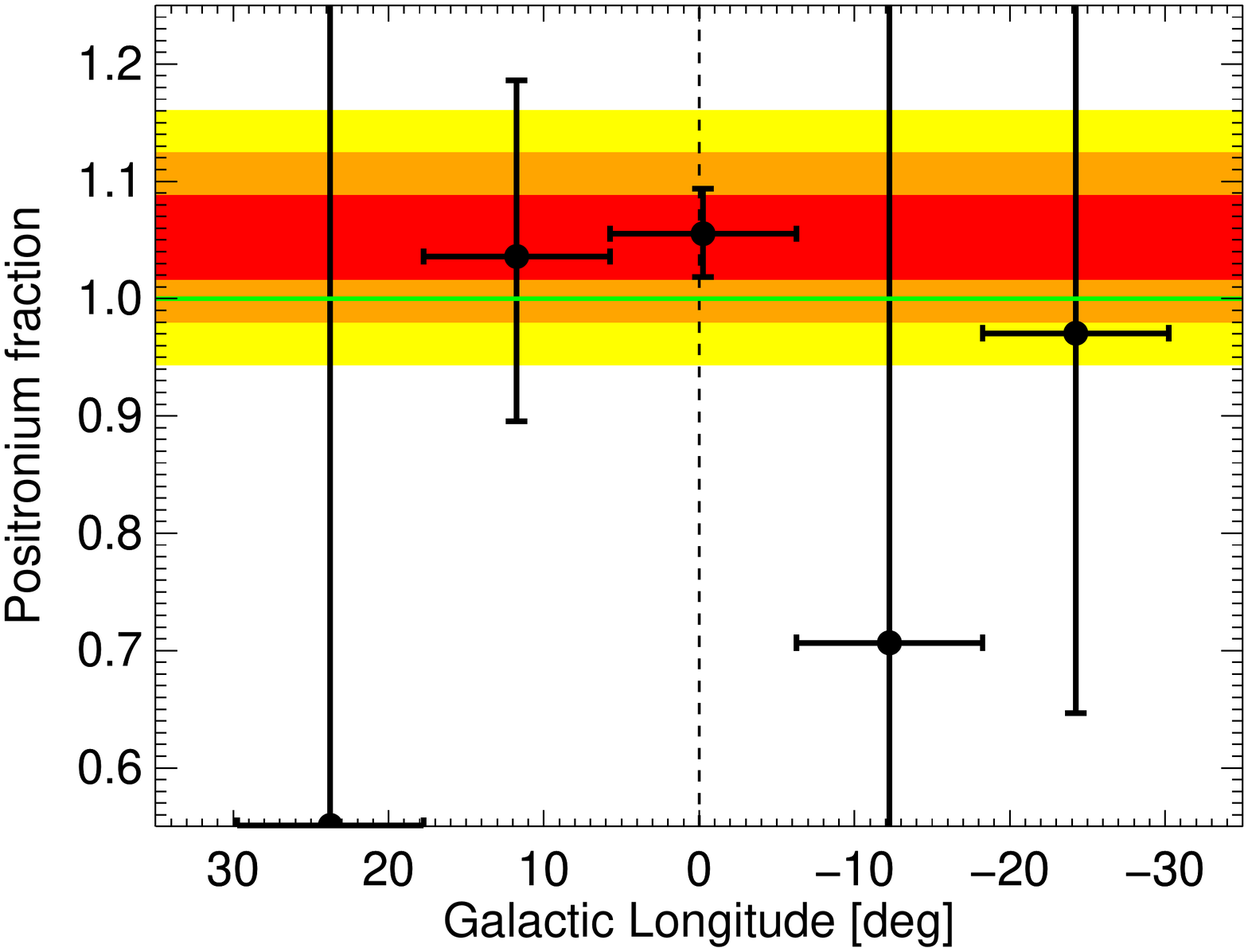}}
	  \cprotect\caption{Derived and fitted spectral parameters (black data points with $1\sigma$ uncertainties) of positron annihilation from simultaneously fitted ROIs. See text for details on how the uncertainty bands (red: $1\sigma$; orange: $2\sigma$; yellow: $3\sigma$) in each parameter are derived.}
\end{figure*}

The spectral parameters, $I_L$, $v_{los}$, $\Gamma_L$, and $f_{Ps}$ are shown as a function of longitude in Figs.~\ref{fig:l_flux_plot} to \ref{fig:l_fps_plot}. In the plots for the FWHM ($\Gamma_L$, Fig.~\ref{fig:l_fwhm_plot}) as well as the Ps fraction ($f_{Ps}$, Fig.~\ref{fig:l_fps_plot}), the weighted mean across the inner $\pm30^{\circ}$ is shown with its 1, 2, and 3$\sigma$ uncertainty as red, orange, and yellow band, respectively. The Doppler-velocity (Fig.~\ref{fig:l_v_plot}) has been fitted by a straight line, here also indicating the model uncertainties by the same colour scheme. For the line flux (Fig.~\ref{fig:l_flux_plot}), varying strongly along the galactic plane, we performed additional spectral fits using the small-ROI sets \verb|S1| to \verb|S3| (see Appendices~\ref{sec:slidingwindow} and \ref{sec:systematics}, using the sliding window method). The signal-to-noise ratios in these spectra are too low to either detect the o-Ps continuum, or constrain the width of the 511~keV line. Thus we constrain the fits in these small ROIs to the values obtained form the larger ROIs: the widths $\Gamma_L$ as well as the ratios $R_{OL}$ are well determined, so that we can interpolate these parameters from the coarse longitude binning to the fine binning. Only the line amplitude (flux) and the Doppler-shift are left as free parameters. In this way, we consecutively determine the line flux in the small ROIs and cross-validate our Doppler measurements. The positronium fraction, $f_{Ps}$, is mostly unconstrained. Within $2\sigma$ uncertainties, all values are consistent with 1.0, and with the fitted value for the entire Galaxy. Constraining the spectral fits subject to $f_{Ps} \leq 1.0$ obtains marginal changes in Doppler-velocities, $\sigma_v \lesssim 20~\%$. In Fig.~\ref{fig:l_flux_plot}, the overlapping small-ROI sets \verb|S1| to \verb|S3| are shown as grey data points, being normalised to the solid angle of such a region, $\Omega_S = \Delta l_S \times \Delta b_S = 3^{\circ} \times 21^{\circ} \times \frac{\pi^2}{32400} \mrm{\frac{sr}{deg^2}} = 1.9\times10^{-2}~\mrm{sr}$. Also here, the uncertainty bands are shown in the same colour scheme, but derived from the small-ROI sets. It can be seen, that the mean values of the small-ROI set directly reflects the derived values of the large-ROI bins, being normalised to the solid angle of a large ROI bin, $\Omega_L = \Delta l_L \times \Delta b_L = 12^{\circ} \times 21^{\circ} \times \frac{\pi^2}{32400} \mrm{\frac{sr}{deg^2}} = 7.8\times10^{-2}~\mrm{sr}$. Thus, our flux estimates are robust. The same is true for the estimated Doppler-velocities in the small ROIs, however with very large uncertainties beyond $-8^{\circ} \lesssim l \lesssim +18^{\circ}$.

\section{Results}\label{sec:results}

\subsection{Galactic rotation}\label{sec:rotation}

We find line-of-sight velocities from $-230_{-150}^{+170}~\mrm{km~s^{-1}}$ (blue side) over $-10\pm40~\mrm{km~s^{-1}}$ (centre) to $-140_{-470}^{+790}~\mrm{km~s^{-1}}$ (red side). This results in an average velocity in the inner radian of $v^{los}_{511} \approx -33 \pm 38~\mrm{km~s^{-1}}$, and a velocity gradient of $\xi_{511} = 4.1~\left(^{+5.5}_{-5.4}\right)_{stat}~\left(^{+0.4}_{-0.5}\right)_{syst}~\mrm{km~s^{-1}~deg^{-1}}$. Here, the systematic uncertainties are estimated from the different ROI sets (Appendix~\ref{sec:slidingwindow}). There is barely evidence for galactic rotation in the 511~keV signal. The orientation of the velocity gradient is aligned with the CO signal, however the spectral uncertainties from the individual ROIs are too large to claim a consistent trend by using this method. For a less conservative estimation of spectral parameters and trends along galactic longitudes, see Appendices~\ref{sec:slidingwindow}, \ref{sec:systematics}, and \ref{sec:systematic_trend}.

Positron annihilation agrees with the galactic CO rotation velocity within $1\sigma$, comparing this value to the velocity gradient in CO longitude-profiles \citep[][$\xi_{CO} \approx 2$--$3~\mrm{km~s^{-1}~deg^{-1}}$]{Dame2001_galvel}. For the $\mathrm{^{26}Al}$ Doppler-shift measurements from \citet{Kretschmer2013_26Al}, a velocity gradient of $\xi_{26Al} \approx 8.5\pm0.9~\mrm{km~s^{-1}~deg^{-1}}$ is estimated in the inner $l\pm30^{\circ}$ ($7.8\pm0.7~\mrm{km~s^{-1}~deg^{-1}}$ using all available data out to $\pm42^{\circ}$). The rotation velocities of 511~keV positron annihilation and 1.8~MeV $\mathrm{^{26}Al}$-decay thus also agree within $1\sigma$, considering the gradient.

\subsection{Annihilation parameters}\label{sec:annihilation_parameters}

The total 511~keV line flux in the inner $\pm 30^{\circ}$ amounts\footnote{The total integrated flux, normalised to the solid angle of $\Omega_{tot} = 60^{\circ} \times 21^{\circ} = 0.39~\mrm{sr}$ amounts to $(3.33^{+0.18}_{-0.15}) \times 10^{-3}~\mrm{ph~cm^{-2}~s^{-1}~sr^{-1}}$.} to $(1.31^{+0.17}_{-0.10}) \times 10^{-3}~\mrm{ph~cm^{-2}~s^{-1}}$, summing the ROI set \verb|L1|, or $(1.20^{+0.05}_{-0.06}) \times 10^{-3}~\mrm{ph~cm^{-2}~s^{-1}}$ for the ROI set \verb|S1|. The fluxes from different binnings are consistent, and also with previous measurements \citep[e.g.][]{Jean2006_511,Churazov2011_511,Siegert2016_511}. From analyses of the sliding window method, alternative ROI sets show enhanced fluxes for positive longitudes ($+5^{\circ} \lesssim l \lesssim +18^{\circ}$), while in the shown ROI set, no flux asymmetry between $+10^{\circ}$ to $+20^{\circ}$ and $-10^{\circ}$ to $-20^{\circ}$ is visible (Fig.~\ref{fig:l_flux_plot}). This is illustrated further in the systematics study (Appendix~\ref{sec:systematic_trend}, Fig.~\ref{fig:l_flux_sys}), and also indicated in the analysis using narrow longitude bins (Fig.~\ref{fig:l_flux_plot}, grey data points and contours). We determine a skewness\footnote{A skewness of zero implies a symmetric distribution about the mean; a negative (positive) skewness means that the left (right) wing of the distribution is longer.} of the longitude-flux-distribution of $-0.15 \pm 0.10$ . The uncertainty on the skewness is estimated by re-sampling the flux values given their uncertainties, which provides other realisations of the same data.

We find a total positron annihilation flux (line plus o-Ps continuum) in the inner radian of $(7.4^{+1.4}_{-1.3}) \times 10^{-3}~\mrm{ph~cm^{-2}~s^{-1}}$. Also here, the integrated flux shows an enhancement at positive longitudes. We illustrate this effect in Appendix~\ref{sec:systematic_trend}. Note that ignoring the o-Ps continuum in the spectral fits, even if the component is not significant, will result in an over-estimation of the line-of-sight velocities by a factor of a few. The o-Ps flux is skewed to $-0.24\pm0.19$.

The weighted average of the astrophysical 511~keV line width, $\Gamma_L$, (Fig.~\ref{fig:l_fwhm_plot}; FWHM above instrumental resolution) along galactic longitudes is $2.43\pm0.14$~keV. This is consistent with previous works \citep[e.g.][finding $2.6\pm0.2$~keV for the bulge region]{Churazov2011_511,Siegert2016_511}. The individual data points coincide within $\lesssim 1.7\sigma$ with the mean. However, the four outer data points are also consistent with larger widths. This might be the same broad line as for the total galactic spectrum (cf. Fig~\ref{fig:gal511spec}), with the far wings of the line drowning in the background. The galactic-wide narrow line width of $2.06\pm0.08$~keV is smaller than the weighted average across the inner radian. This may either mean that different line-of-sights are dominated by different annihilation conditions, or that the superposition of broad and narrow line features in the low signal-to-noise spectra smear out to a single broader line. From the sliding window method, we determine a skewness of $-0.34\pm0.07$ for the FWHM.

From the line and the three-photon continuum flux, the Ps fraction, $f_{Ps}$, can be derived (see Appendix \ref{sec:appendix_functions}). This derived value is a weak function of the flux ratio $R_{OL}$, and holds large uncertainties, depending on how the spectrum is modelled. As a function of longitude, $f_{Ps}$ is constant, consistent with the galactic mean, cf. Fig.~\ref{fig:l_fps_plot}, and also consistent with the physical limit of 1.0. Fitted $f_{Ps}$-values greater than 1.0 are unphysical, and may occur in the energy range used here, because the low-energy part of the spectrum is too narrow to constrain the Ps fraction, when the signal-to-noise ratio is low. Performing the spectral fits subject to $f_{Ps} \leq 1.0$ provides the same results in the individual components. The fitted Doppler-shifts vary by at most $0.2\sigma$ when the fits are constrained. Additional systematic uncertainties for the Doppler-shifts are about $10$--$20~\mrm{km~s^{-1}}$. Considering the uncertainties for large $|l|$, the Ps fraction could be much lower in the disk part than in the bulge part. As the values stay the same with longitude, the line flux and the o-Ps flux must vary in the same manner. Thus, if one positron annihilation flux component shows a systematic trend, the other flux component must show the same trend, as seen for the asymmetry in the line and o-Ps measurements from the sliding window method (see also Appendix \ref{sec:systematic_trend}).

\section{Discussion and summary}

\subsection{Discussion}\label{sec:discussion}

The l-v-curve of the 511~keV line in the central radian of the Galaxy is

\begin{enumerate}
	\item consistent with galactic rotation velocities from CO: $|v_{\mrm{los}}^{511}| \approx |v_{\mrm{los}}^{\mrm{CO}}| \lesssim 220~\mrm{km~s^{-1}}$.
	\item consistent with Doppler-velocities from the positron source $\mrm{^{26}Al}$ or faster: $|v_{\mrm{los}}^{511}| \gtrsim |v_{\mrm{los}}^{\mrm{^{26}Al}}| \lesssim 300~\mrm{km~s^{-1}}$.
	\item consistent with zero (flat velocity curve, no gradient along longitudes): $|v_{\mrm{los}}^{511}| \approx 0~\mrm{km~s^{-1}}$.
\end{enumerate}

The measured velocity gradient of $\xi_{511} = 4.1\pm5.9~\mrm{km~s^{-1}~deg^{-1}}$ in the inner radian of the Milky Way has a direct impact on the line broadening of the galaxy-wide 511~keV line: The $2\sigma$ upper limit on $\xi_{511}$ of $\lesssim 12~\mrm{km~s^{-1}~deg^{-1}}$ can result in a line broadening of up to $0.25~\mrm{keV}$ across the angular range of one $12^{\circ}$-ROI, and up to $1.25~\mrm{keV}$ for the entire inner radian. From galactic rotation velocities as measured in CO, only 0.05~keV of the 511~keV line width would originate from Doppler broadening of galactic kinematics in such a ROI (0.26~keV for the inner $60^{\circ}$). This means that the annihilation conditions could be different to what has been inferred in previous studies, not considering the galactic kinematics. Doppler broadening adds in quadrature, so that the FWHM of the 511~keV line in the bulge may be narrower than 2~keV. As a result, the allowed ISM temperatures and ionisation states may cover a different area in their combined data space \citep[e.g.][]{Churazov2005_511,Churazov2011_511}. Moreover, alternative positron annihilation branches, such as on dust grains \citep{Guessoum2005_511} or polycyclic aromatic hydrocarbon molecules \citep{Guessoum2010_511} may be more important than considered previously.

\subsection{Summary}\label{sec:summary}

In this paper, we reported on a kinematic analysis of the galactic 511~keV line from positron annihilation in different regions of the Milky Way. The main finding are summarised as follows:

\begin{enumerate}
	\item The measured 511~keV Doppler-velocities, between $\approx -400$ and $\approx +650~\mrm{km~s^{-1}}$, are aligned with the orientation of galactic rotation, however with large uncertainties which are consistent with line-of-sight velocities from CO, $\mrm{^{26}Al}$, and also with zero.
	\item Through our kinematic study, we find evidence for positrons being slowed down from initially larger (relativistic) velocities to average kinetic energies below a few tens of eV, and thus confirm previous studies.
	\item From the measured velocity gradient of $\lesssim 12~\mrm{km~s^{-1}~deg^{-1}}$, we derive an upper limit on additional Doppler-broadening from galactic kinematics of $<1.25$~keV for the inner radian ($<0.25$~keV for a $12^{\circ}$ region), possibly changing the true annihilation conditions.
\end{enumerate}

\begin{acknowledgements}
This research was supported by the German DFG cluster of excellence `Origin and Structure of the Universe'. The INTEGRAL/SPI project has been completed under the responsibility and leadership of CNES; we are grateful to ASI, CEA, CNES, DLR, ESA, INTA, NASA and OSTC for support of this ESA space science mission. TS is supported by the German Research Society (DFG-Forschungsstipendium SI 2502/1-1). FHP is supported by an Australian Government Research Training Program (RTP) Scholarship and the Alex Rogers Travelling Scholarship.
\end{acknowledgements}

\bibliographystyle{aa} 
\bibliography{alles} 

\begin{thebibliography}{55}
\expandafter\ifx\csname natexlab\endcsname\relax\def\natexlab#1{#1}\fi

\bibitem[{{Aharonian} \& {Atoyan}(1981)}]{Aharonian1981_TPA}
{Aharonian}, F.~A. \& {Atoyan}, A.~M. 1981, Physics Letters B, 99, 301

\bibitem[{{Andrae} {et~al.}(2010){Andrae}, {Schulze-Hartung}, \&
  {Melchior}}]{Andrae2010_chi2}
{Andrae}, R., {Schulze-Hartung}, T., \& {Melchior}, P. 2010, ArXiv e-prints,
  1012.3754

\bibitem[{{Atti{\'e}} {et~al.}(2003){Atti{\'e}}, {Cordier}, {Gros}, {Laurent},
  {Schanne}, {Tauzin}, {von Ballmoos}, {Bouchet}, {Jean}, {Kn{\"o}dlseder},
  {Mandrou}, {Paul}, {Roques}, {Skinner}, {Vedrenne}, {Georgii}, {von Kienlin},
  {Lichti}, {Sch{\"o}nfelder}, {Strong}, {Wunderer}, {Shrader}, {Sturner},
  {Teegarden}, {Weidenspointner}, {Kiener}, {Porquet}, {Tatischeff}, {Crespin},
  {Joly}, {Andr{\'e}}, {Sanchez}, \& {Leleux}}]{Attie2003_SPI}
{Atti{\'e}}, D., {Cordier}, B., {Gros}, M., {et~al.} 2003, \aap, 411, L71

\bibitem[{{Baade}(1946)}]{Baade1946_RRLgalcen}
{Baade}, W. 1946, \pasp, 58, 249

\bibitem[{{Beacom} \& {Y{\"u}ksel}(2006)}]{Beacom2006_511}
{Beacom}, J.~F. \& {Y{\"u}ksel}, H. 2006, Physical Review Letters, 97, 071102

\bibitem[{{Bisnovatyi-Kogan} \& {Pozanenko}(2017)}]{Bisnovatyi-Kogan2017_511}
{Bisnovatyi-Kogan}, G.~S. \& {Pozanenko}, A.~S. 2017, Astrophysics, 60, 223

\bibitem[{{Bouchet} {et~al.}(2010{\natexlab{a}}){Bouchet}, {Roques}, \&
  {Jourdain}}]{Bouchet2010_positron}
{Bouchet}, L., {Roques}, J.~P., \& {Jourdain}, E. 2010{\natexlab{a}}, \apj,
  720, 1772

\bibitem[{{Bouchet} {et~al.}(2010{\natexlab{b}}){Bouchet}, {Roques}, \&
  {Jourdain}}]{Bouchet2010_511}
{Bouchet}, L., {Roques}, J.~P., \& {Jourdain}, E. 2010{\natexlab{b}}, \apj,
  720, 1772

\bibitem[{{Cash}(1979)}]{Cash1979_cstat}
{Cash}, W. 1979, \apj, 228, 939

\bibitem[{{Churazov} {et~al.}(2011){Churazov}, {Sazonov}, {Tsygankov},
  {Sunyaev}, \& {Varshalovich}}]{Churazov2011_511}
{Churazov}, E., {Sazonov}, S., {Tsygankov}, S., {Sunyaev}, R., \&
  {Varshalovich}, D. 2011, \mnras, 411, 1727

\bibitem[{{Churazov} {et~al.}(2005){Churazov}, {Sunyaev}, {Sazonov},
  {Revnivtsev}, \& {Varshalovich}}]{Churazov2005_511}
{Churazov}, E., {Sunyaev}, R., {Sazonov}, S., {Revnivtsev}, M., \&
  {Varshalovich}, D. 2005, \mnras, 357, 1377

\bibitem[{{Clement} {et~al.}(2001){Clement}, {Muzzin}, {Dufton}, {Ponnampalam},
  {Wang}, {Burford}, {Richardson}, {Rosebery}, {Rowe}, \&
  {Hogg}}]{Clement2001_RRL}
{Clement}, C.~M., {Muzzin}, A., {Dufton}, Q., {et~al.} 2001, \aj, 122, 2587

\bibitem[{{Clementini} {et~al.}(2001){Clementini}, {Federici}, {Corsi},
  {Cacciari}, {Bellazzini}, \& {Smith}}]{Clementini2001_RRL}
{Clementini}, G., {Federici}, L., {Corsi}, C., {et~al.} 2001, \apjl, 559, L109

\bibitem[{{Crocker} {et~al.}(2017){Crocker}, {Ruiter}, {Seitenzahl}, {Panther},
  {Sim}, {Baumgardt}, {M{\"o}ller}, {Nataf}, {Ferrario}, {Eldridge}, {White},
  {Tucker}, \& {Aharonian}}]{Crocker2017_511_91bg}
{Crocker}, R.~M., {Ruiter}, A.~J., {Seitenzahl}, I.~R., {et~al.} 2017, Nature
  Astronomy, 1, 0135

\bibitem[{{Dame} {et~al.}(2001){Dame}, {Hartmann}, \&
  {Thaddeus}}]{Dame2001_galvel}
{Dame}, T.~M., {Hartmann}, D., \& {Thaddeus}, P. 2001, \apj, 547, 792

\bibitem[{{Dan} {et~al.}(2015){Dan}, {Guillochon}, {Br{\"u}ggen},
  {Ramirez-Ruiz}, \& {Rosswog}}]{Dan2015_SNIa_merger}
{Dan}, M., {Guillochon}, J., {Br{\"u}ggen}, M., {Ramirez-Ruiz}, E., \&
  {Rosswog}, S. 2015, \mnras, 454, 4411

\bibitem[{{Diehl} {et~al.}(2010){Diehl}, {Lang}, {Martin}, {Ohlendorf},
  {Preibisch}, {Voss}, {Jean}, {Roques}, {von Ballmoos}, \&
  {Wang}}]{Diehl2010_ScoCen}
{Diehl}, R., {Lang}, M.~G., {Martin}, P., {et~al.} 2010, \aap, 522, A51

\bibitem[{{Diehl} {et~al.}(2018){Diehl}, {Siegert}, {Greiner}, {Krause},
  {Kretschmer}, {Lang}, {Pleintinger}, {Strong}, {Weinberger}, \&
  {Zhang}}]{Diehl2018_SPI}
{Diehl}, R., {Siegert}, T., {Greiner}, J., {et~al.} 2018, \aap, 611, A12

\bibitem[{{Ellis} \& {Bland-Hawthorn}(2018)}]{Ellis2018_leptonium}
{Ellis}, S.~C. \& {Bland-Hawthorn}, J. 2018, European Physical Journal D, 72,
  18

\bibitem[{{Ellison} {et~al.}(1990){Ellison}, {Jones}, \&
  {Ramaty}}]{Ellison1990_positrons}
{Ellison}, C.~D., {Jones}, C.~F., \& {Ramaty}, R. 1990, International Cosmic
  Ray Conference, 4, 68

\bibitem[{{Endt}(1990)}]{Endt1990_nuclei21-44}
{Endt}, P.~M. 1990, Nuclear Physics A, 521, 1

\bibitem[{Gryzi\ifmmode~\acute{n}\else
  \'{n}\fi{}ski(1965)}]{Gryzinski1965_particle_collisions}
Gryzi\ifmmode~\acute{n}\else \'{n}\fi{}ski, M. 1965, Phys. Rev., 138, A305

\bibitem[{{Guessoum} {et~al.}(2005){Guessoum}, {Jean}, \&
  {Gillard}}]{Guessoum2005_511}
{Guessoum}, N., {Jean}, P., \& {Gillard}, W. 2005, \aap, 436, 171

\bibitem[{{Guessoum} {et~al.}(2010){Guessoum}, {Jean}, \&
  {Gillard}}]{Guessoum2010_511}
{Guessoum}, N., {Jean}, P., \& {Gillard}, W. 2010, \mnras, 402, 1171

\bibitem[{Hastings(1970)}]{Hastings1970_MH_algorithm}
Hastings, W.~K. 1970, Biometrika, 57, 97

\bibitem[{{Haymes} {et~al.}(1969){Haymes}, {Ellis}, {Fishman}, {Glenn}, \&
  {Kurfess}}]{Haymes1969_511}
{Haymes}, R.~C., {Ellis}, D.~V., {Fishman}, G.~J., {Glenn}, S.~W., \&
  {Kurfess}, J.~D. 1969, \apj, 157, 1455

\bibitem[{{Jean} {et~al.}(2006){Jean}, {Kn{\"o}dlseder}, {Gillard}, {Guessoum},
  {Ferri{\`e}re}, {Marcowith}, {Lonjou}, \& {Roques}}]{Jean2006_511}
{Jean}, P., {Kn{\"o}dlseder}, J., {Gillard}, W., {et~al.} 2006, \aap, 445, 579

\bibitem[{{Jeffers} {et~al.}(2012){Jeffers}, {Min}, {Waters}, {Canovas},
  {Rodenhuis}, {de Juan Ovelar}, {Chies-Santos}, \&
  {Keller}}]{Jeffers2012_RCorBor_dust}
{Jeffers}, S.~V., {Min}, M., {Waters}, L.~B.~F.~M., {et~al.} 2012, \aap, 539,
  A56

\bibitem[{{Johnson} {et~al.}(1972){Johnson}, {Harnden}, \&
  {Haymes}}]{Johnson1972_511}
{Johnson}, III, W.~N., {Harnden}, Jr., F.~R., \& {Haymes}, R.~C. 1972, \apjl,
  172, L1

\bibitem[{{Kinzer} {et~al.}(2001){Kinzer}, {Milne}, {Kurfess}, {Strickman},
  {Johnson}, \& {Purcell}}]{Kinzer2001_511}
{Kinzer}, R.~L., {Milne}, P.~A., {Kurfess}, J.~D., {et~al.} 2001, \apj, 559,
  282

\bibitem[{Kn{\"o}dlseder(1997)}]{Knoedlseder1997_PhD}
Kn{\"o}dlseder, J. 1997, Theses, {Universit{\'e} Paul Sabatier - Toulouse III}

\bibitem[{{Kn{\"o}dlseder} {et~al.}(2005){Kn{\"o}dlseder}, {Jean}, {Lonjou},
  {Weidenspointner}, {Guessoum}, {Gillard}, {Skinner}, {von Ballmoos},
  {Vedrenne}, {Roques}, {Schanne}, {Teegarden}, {Sch{\"o}nfelder}, \&
  {Winkler}}]{Knoedlseder2005_511}
{Kn{\"o}dlseder}, J., {Jean}, P., {Lonjou}, V., {et~al.} 2005, \aap, 441, 513

\bibitem[{{{Kretschmer}}(2011)}]{Kretschmer2011_PhD}
{{Kretschmer}}. 2011, Dissertation, {Technische Universit\"at M\"unchen},
  {M\"unchen}

\bibitem[{{Kretschmer} {et~al.}(2013){Kretschmer}, {Diehl}, {Krause},
  {Burkert}, {Fierlinger}, {Gerhard}, {Greiner}, \&
  {Wang}}]{Kretschmer2013_26Al}
{Kretschmer}, K., {Diehl}, R., {Krause}, M., {et~al.} 2013, \aap, 559, A99

\bibitem[{{Kunder} {et~al.}(2016){Kunder}, {Rich}, {Koch}, {Storm}, {Nataf},
  {De Propris}, {Walker}, {Bono}, {Johnson}, {Shen}, \&
  {Li}}]{Kunder2016_lvstars}
{Kunder}, A., {Rich}, R.~M., {Koch}, A., {et~al.} 2016, \apjl, 821, L25

\bibitem[{{Lallement} {et~al.}(2014){Lallement}, {Vergely}, {Valette},
  {Puspitarini}, {Eyer}, \& {Casagrande}}]{Lallement2014_OB}
{Lallement}, R., {Vergely}, J.-L., {Valette}, B., {et~al.} 2014, \aap, 561, A91

\bibitem[{{Martinez-Valpuesta} \&
  {Gerhard}(2011)}]{Martinez-Valpuesta2011_boxybulgebar}
{Martinez-Valpuesta}, I. \& {Gerhard}, O. 2011, \apjl, 734, L20

\bibitem[{{Metropolis} {et~al.}(1953){Metropolis}, {Rosenbluth}, {Rosenbluth},
  {Teller}, \& {Teller}}]{Metropolis1953_MH_algorithm}
{Metropolis}, N., {Rosenbluth}, A.~W., {Rosenbluth}, M.~N., {Teller}, A.~H., \&
  {Teller}, E. 1953, \jcp, 21, 1087

\bibitem[{{Milne} \& {Leising}(1997)}]{Milne1997_511}
{Milne}, P.~A. \& {Leising}, M.~D. 1997, in American Institute of Physics
  Conference Series, Vol. 410, Proceedings of the Fourth Compton Symposium, ed.
  C.~D. {Dermer}, M.~S. {Strickman}, \& J.~D. {Kurfess}, 1017--1021

\bibitem[{{Murphy} {et~al.}(2005){Murphy}, {Share}, {Skibo}, \&
  {Kozlovsky}}]{Murphy2005_posiloss}
{Murphy}, R.~J., {Share}, G.~H., {Skibo}, J.~G., \& {Kozlovsky}, B. 2005,
  \apjs, 161, 495

\bibitem[{{Ore} \& {Powell}(1949)}]{Ore1949_511}
{Ore}, A. \& {Powell}, J.~L. 1949, Physical Review, 75, 1696

\bibitem[{{Pakmor} {et~al.}(2013){Pakmor}, {Kromer}, {Taubenberger}, \&
  {Springel}}]{Pakmor2013_SNIa_merger}
{Pakmor}, R., {Kromer}, M., {Taubenberger}, S., \& {Springel}, V. 2013, \apjl,
  770, L8

\bibitem[{{Perets} {et~al.}(2010){Perets}, {Gal-Yam}, {Mazzali}, {Arnett},
  {Kagan}, {Filippenko}, {Li}, {Arcavi}, {Cenko}, {Fox}, {Leonard}, {Moon},
  {Sand}, {Soderberg}, {Anderson}, {James}, {Foley}, {Ganeshalingam}, {Ofek},
  {Bildsten}, {Nelemans}, {Shen}, {Weinberg}, {Metzger}, {Piro}, {Quataert},
  {Kiewe}, \& {Poznanski}}]{Perets2010_SN2005E_44Ti}
{Perets}, H.~B., {Gal-Yam}, A., {Mazzali}, P.~A., {et~al.} 2010, \nat, 465, 322

\bibitem[{{Purcell} {et~al.}(1997){Purcell}, {Cheng}, {Dixon}, {Kinzer},
  {Kurfess}, {Leventhal}, {Saunders}, {Skibo}, {Smith}, \&
  {Tueller}}]{Purcell1997_511}
{Purcell}, W.~R., {Cheng}, L.-X., {Dixon}, D.~D., {et~al.} 1997, \apj, 491, 725

\bibitem[{Siegert(2017)}]{Siegert2017_PhD}
Siegert, T. 2017, Dissertation, Technische Universit{\"a}t M{\"u}nchen,
  Published online at https://mediatum.ub.tum.de/node?id=1340342

\bibitem[{{Siegert} {et~al.}(2016){Siegert}, {Diehl}, {Khachatryan}, {Krause},
  {Guglielmetti}, {Greiner}, {Strong}, \& {Zhang}}]{Siegert2016_511}
{Siegert}, T., {Diehl}, R., {Khachatryan}, G., {et~al.} 2016, \aap, 586, A84

\bibitem[{{Siegert} {et~al.}(2019){Siegert}, {Diehl}, {Weinberger},
  {Pleintinger}, {Greiner}, \& {Zhang}}]{Siegert2019_SPI}
{Siegert}, T., {Diehl}, R., {Weinberger}, C., {et~al.} 2019, arXiv e-prints

\bibitem[{{Skinner} {et~al.}(2014){Skinner}, {Diehl}, {Zhang}, {Bouchet}, \&
  {Jean}}]{Skinner2014_511}
{Skinner}, G., {Diehl}, R., {Zhang}, X., {Bouchet}, L., \& {Jean}, P. 2014, in
  Proceedings of the 10th INTEGRAL Workshop: ''A Synergistic View of the
  High-Energy Sky'' (INTEGRAL 2014). 15-19 September 2014. Annapolis, MD, USA.
  Published online at http://pos.sissa.it/cgi-bin/reader/conf.cgi?confid=228,
  id.054, 054

\bibitem[{{Svensson}(1982)}]{Svensson1982_ann_spec}
{Svensson}, R. 1982, \apj, 258, 321

\bibitem[{{Vedrenne} {et~al.}(2003){Vedrenne}, {Roques}, {Sch{\"o}nfelder},
  {Mandrou}, {Lichti}, {von Kienlin}, {Cordier}, {Schanne}, {Kn{\"o}dlseder},
  {Skinner}, {Jean}, {Sanchez}, {Caraveo}, {Teegarden}, {von Ballmoos},
  {Bouchet}, {Paul}, {Matteson}, {Boggs}, {Wunderer}, {Leleux},
  {Weidenspointner}, {Durouchoux}, {Diehl}, {Strong}, {Cass{\'e}}, {Clair}, \&
  {Andr{\'e}}}]{Vedrenne2003_SPI}
{Vedrenne}, G., {Roques}, J.-P., {Sch{\"o}nfelder}, V., {et~al.} 2003, \aap,
  411, L63

\bibitem[{{Waldman} {et~al.}(2011){Waldman}, {Sauer}, {Livne}, {Perets},
  {Glasner}, {Mazzali}, {Truran}, \& {Gal-Yam}}]{Waldman2011_SN2005E_44Ti}
{Waldman}, R., {Sauer}, D., {Livne}, E., {et~al.} 2011, \apj, 738, 21

\bibitem[{{Wegg} \& {Gerhard}(2013)}]{Wegg2013_bulgebar}
{Wegg}, C. \& {Gerhard}, O. 2013, \mnras, 435, 1874

\bibitem[{{Weidenspointner} {et~al.}(2008){Weidenspointner}, {Skinner}, {Jean},
  {Kn{\"o}dlseder}, {von Ballmoos}, {Diehl}, {Strong}, {Cordier}, {Schanne}, \&
  {Winkler}}]{Weidenspointner2008_511b}
{Weidenspointner}, G., {Skinner}, G.~K., {Jean}, P., {et~al.} 2008, \nar, 52,
  454

\bibitem[{{Winkler} {et~al.}(2003){Winkler}, {Courvoisier}, {Di Cocco},
  {Gehrels}, {Gim{\'e}nez}, {Grebenev}, {Hermsen}, {Mas-Hesse}, {Lebrun},
  {Lund}, {Palumbo}, {Paul}, {Roques}, {Schnopper}, {Sch{\"o}nfelder},
  {Sunyaev}, {Teegarden}, {Ubertini}, {Vedrenne}, \&
  {Dean}}]{Winkler2003_INTEGRAL}
{Winkler}, C., {Courvoisier}, T.~J.-L., {Di Cocco}, G., {et~al.} 2003, \aap,
  411, L1

\bibitem[{{Zhang} {et~al.}(2014){Zhang}, {Jeffery}, {Chen}, \&
  {Han}}]{Zhang2014_RCorBor_WDmerger}
{Zhang}, X., {Jeffery}, C.~S., {Chen}, X., \& {Han}, Z. 2014, \mnras, 445, 660

\end{thebibliography}

\appendix

\section{Sliding window method}\label{sec:slidingwindow}

The "sliding window method" was first introduced by \citet{Kretschmer2011_PhD} and \citet{Kretschmer2013_26Al} for measuring the kinematics of $\mrm{^{26}Al}$ emission at 1.8~MeV. The method described here aims to improve the sensitivity for Doppler-shift measurements in the 511~keV energy region with SPI. The major principle is to scan the galactic plane by using the sky model, and splitting it in only two complementary parts. One model defines the ROI again, and each ROI has its own complement,

\begin{equation}
	M_{COMP}^r(l,b) = A_{COMP}^r \times \left(M_{511}(l,b) - M_{ROI}^r(l,b)\right)\mrm{.}	
\label{eq:complementary_map_roi}
\end{equation}

Also here, a complete description of the full sky is necessary to account for the field of view of SPI. The exact morphology of the ROIs has marginal influence on the spectral results. This is verified, for example, by ignoring the galactic centre source (GCS) in the central ROI, and also by various tests considering smaller longitude bins, as well as latitudinal changes, and comparisons to the base-line method (see Appendices~\ref{sec:method_comparison} and \ref{sec:systematics}). This method is less correlated, as the number of fitted parameters is reduced from prior knowledge on the imaging part of the analysis \citep{Kretschmer2011_PhD}. Performing the same analysis without the GCS, results in the same Doppler-shift values within less than $10~\mrm{km~s^{-1}}$.

The process described above is repeated by varying $l_0^r$ to scan the galactic plane. For further consistency checks and to search for trends in the line flux, we perform a high angular resolution analysis with $\Delta l = 3^{\circ}$ ROI bins. The latitudinal extent of the galactic 511~keV emission is the least well-determined parameter. We therefore also dissect the central $\pm 15^{\circ}$ into 21 ROIs of $\Delta l \times \Delta b = 10^{\circ} \times 3^{\circ}$ (see also Appendix~\ref{sec:systematics}). This allows us to validate the results obtained from the longitude investigation, and provides additional information as a function of latitude.

In this way, individual spectra for different regions of the sky are extracted. The spectra per ROI are determined sequentially with only two input images at any step, i.e. the ROI and the complementary map, zeroing the ROI part, cf. Eqs.~(\ref{eq:ROI_math_definition}) and (\ref{eq:complementary_map_roi}). This fitting procedure is repeated for all ROIs (and complementary maps) that we define in our analysis.

Any ROI segment covers many emission regions at different distances. Nearby sources have a larger solid angle at fixed physical scale, and might only be partially included, whereas distant annihilation sites might be completely covered but with reduced intensity in small ROIs. It has been shown \citep{Skinner2014_511,Siegert2016_511} that centroid shifts of 2D-Gaussian model templates of the order of $1^{\circ}$ and less are identifiable for the strong 511~keV emission with SPI. Therefore, we try to trace gradual variations in the flux profile of positron annihilation by using smaller ROIs with $\Delta l = 3^{\circ}$. If there was enhanced or reduced emission in different lines-of-sight due to slight variations in the true galactic emission with respect to the smooth model, we would capture these in the narrow bins. In particular, we create four different ROI sets for the large-longitude bins ($\Delta l = 12^{\circ}$; named \verb|L1|, \verb|L2|, \verb|L3|, and \verb|L4|), and three ROI sets for the small-longitude bins ($\Delta l = 3^{\circ}$; named \verb|S1|, \verb|S2|, and \verb|S3|). The ROI set \verb|L1| ranges from $l=-29.75^{\circ}$ to $l=+30.25^{\circ}$, in steps of $\Delta l = 12^{\circ}$ bins, for a total of five ROIs. The other large-ROI sets \verb|L2| (\verb|L3|, \verb|L4|) are shifted in $l_0$ with respect to \verb|L1| by $3^{\circ}$ ($6^{\circ}$, $9^{\circ}$). For the small bin widths, the ROIs of \verb|S1| range from $l=-32.75^{\circ}$ to $l=+34.25^{\circ}$, in steps of $\Delta l = 3^{\circ}$ bins, for a total of 22 independent ROIs. Also here, the alternative choices of the binning, \verb|S2| and \verb|S3|, are shifted in $l_0$ by $1^{\circ}$ and $2^{\circ}$, respectively.

\section{Base-line method for spectral extraction and comparisons}\label{sec:method_comparison}

In Sect.~\ref{sec:simultan_fit}, we introduced the most unbiased but least sensitive method to extract spectral information from different regions in the sky. Here, we show the results from this analysis, and compare them to those from the sliding window method, Appendix~\ref{sec:slidingwindow}. The difference between the methods are the constraints on the imaging information: while a blind search for spectral variability may also assume ignorance of the emission morphology, an already understood image on large scale can be used to exploit as much spectral information as possible. Likewise, the instrumental $\gamma$-ray background (BG) could also be pre-determined from a large-scale fit, and then fixed, to only allow the ROIs to vary within the data-given variations. We name these three approaches for the data analysis:
\begin{itemize}
	\item "Free BG", according to Sect.~\ref{sec:simultan_fit} in a simultaneous fit of all ROIs plus the remaining Galaxy
	\item "Fixed BG", again performing a simultaneous fit, but with fixed background from the all-sky model fit, Fig.~\ref{fig:gal511spec}
	\item "Sliding Window", according to Appendix~\ref{sec:slidingwindow}, using only two input map - the ROI and its complement - evaluating the ROIs in sequence
\end{itemize}

\begin{figure}[!ht]%
\centering
\includegraphics[width=0.9\columnwidth,trim=0.74in 1.04in 1.49in 0.44in,clip=true]{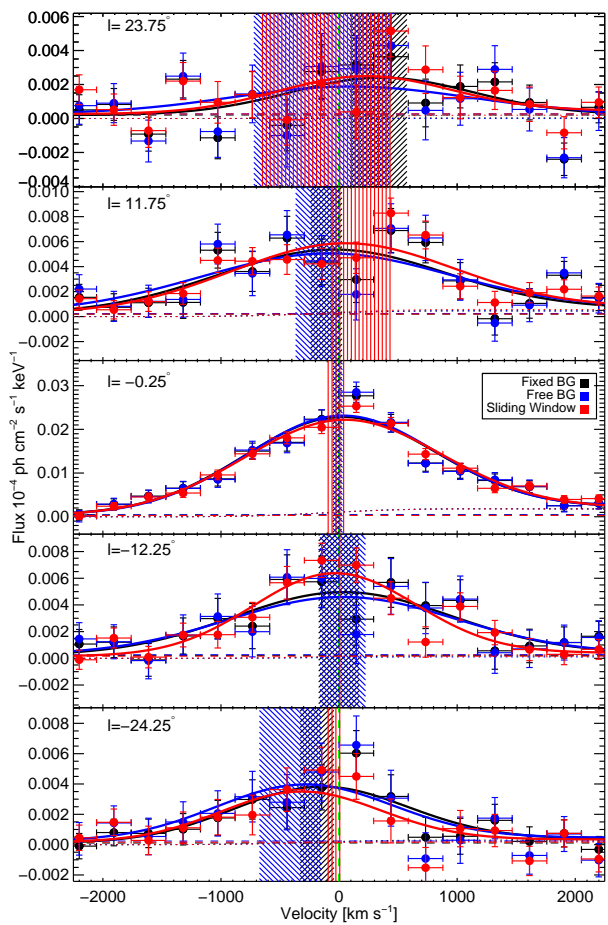}%
\cprotect\caption{Extracted ROI spectra ($1\sigma$ errors bars) in velocity space for three analysis methods. Each spectrum has been fitted with the spectral model of Eq.~(\ref{eq:fit_fun_total}, thick solid lines), similar to Fig.~\ref{fig:specs_L1}. The fitted Doppler-shifts and their uncertainties are marked by the hatched areas in the respective colours. The spectra are consistent with each other, and their fitted Doppler-shifts coincide.}
\label{fig:three_methods_specs}%
\end{figure}

The resulting spectra from the three methods described above are shown in Fig.~\ref{fig:three_methods_specs}. The spectra are consistent, and the uncertainties on the individual data points from the "free BG" and "fixed BG" case are between 1 to 2 times larger than for the "sliding window method".

The base-line method from Sect.~\ref{sec:simultan_fit} tends to ask too many questions to too little data in the correlated simultaneous fit of five small ROIs. As we use the same data set and emission model as \citet{Siegert2016_511}, already describing the SPI data correctly, we can set imaging constraints to allow for more spectral variability, with less parameters, which possibly over-parametrise the model. Therefore, the "sliding window method" can be less correlated in individual cases.

From the sliding window method, we estimate the significance of galactic rotation in 511~keV to $1.6$--$2.0\sigma$. This is done by comparing two empirical models describing the velocity distribution along galactic longitudes. By fitting a straight line to the data (cf. Figs.~\ref{fig:lvdiagram} and \ref{fig:l_v_plot}), we estimate the velocity gradient and bulk velocity, and obtain a measure for the goodness of fit, $\chi^2_{rot}$. Alternatively, we fit a constant to the data points to obtain a fit quality measure for the null-hypothesis, i.e. no rotation, $\chi^2_{flat}$. As these models are nested, we can perform a Pearson $\chi^2$-test with $\Delta \chi^2 = \chi^2_{flat} - \chi^2_{rot}$ and one degree of freedom difference. As we define four different, partially overlapping ROI sets, we estimate our systematic uncertainties in this way, and provide individual results in Tab.~\ref{tab:lvchi2}.

\begin{table}[!ht]%
\centering
\begin{tabular}{l|rrrrrr}
\hline
 & $\chi^2_{flat}~(\nu)$ & $\chi^2_{rot}~(\nu)$ & $\Delta \chi^2$ & $\sigma$ & $v_{bulk}$ & $\xi_{511}$\\
\hline
\verb|L1| & 5.4 (4)                   & 2.7 (3)        & 2.7    & 1.6   & $-6_{-22}^{+21}$  & $+6.0_{-3.4}^{+3.4}$ \\
\verb|L2| & 5.7 (4)                   & 2.4 (3)        & 3.3    & 1.8   & $+6_{-24}^{+25}$  & $+6.8_{-3.7}^{+3.6}$ \\
\verb|L3| & 9.0 (5)                   & 5.7 (4)        & 3.3    & 1.8   & $+10_{-25}^{+25}$  & $+5.3_{-2.8}^{+2.7}$ \\
\verb|L4| & 7.0 (4)                   & 3.0 (3)        & 4.0    & 2.0   & $-27_{-26}^{+25}$  & $+5.8_{-2.8}^{+2.8}$ \\
\hline
\end{tabular}
\caption{Model comparisons between a flat ($\chi^2_{flat}$) and an ascending ($\chi^2_{rot}$) longitude-velocity curve for the four different ROI sets. The $\chi^2$-values have been determined by a fit to the l-v-diagram, Fig.~\ref{fig:lvdiagram}, with $\nu$ degrees of freedom. By comparing the fitted values in a $\chi^2$-test ($\Delta \chi^2$-values), the significances in units of $\sigma$ are derived. In the last two columns, the best fit parameters of an ascending curve are provided, with $v_{bulk}$ as the intersect at $l=0^{\circ}$ in units of $\mrm{km~s^{-1}}$, and $\xi_{511}$ as the velocity gradient in units of $\mrm{km~s^{-1}~deg^{-1}}$.}
\label{tab:lvchi2}
\end{table}

\begin{figure*}[!ht]
	\centering
		  \subfloat[Integrated 511~keV line flux $I_L$. \label{fig:compare_flux_plot}]{\includegraphics[width=0.89\columnwidth,trim=0.68in 0.72in 0.97in 0.96in,clip=true]{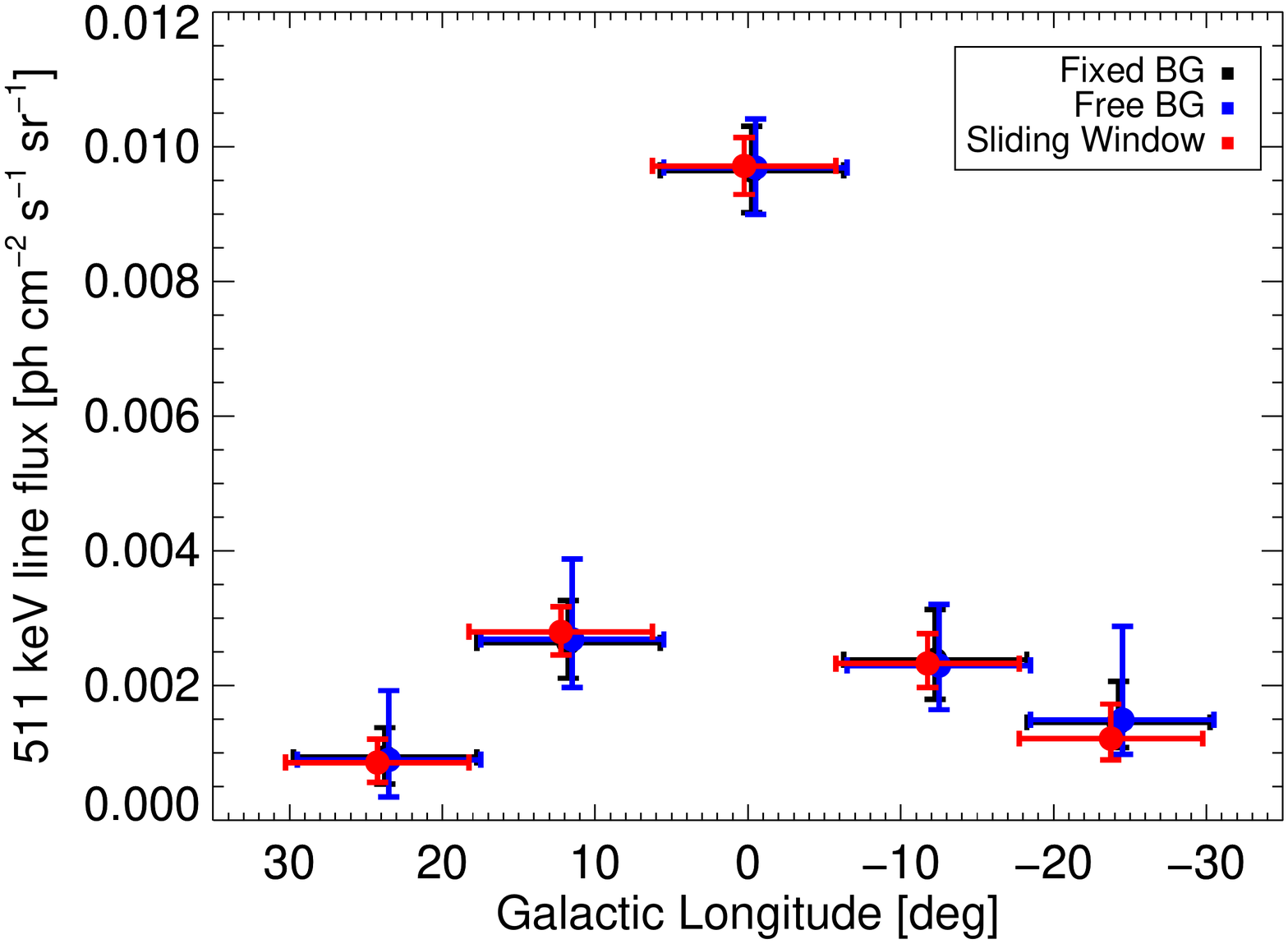}}~
			\subfloat[Line of sight Doppler-velocity $v_{los}$. \label{fig:compare_v_plot}]{\includegraphics[width=0.89\columnwidth,trim=0.68in 0.72in 0.97in 0.96in,clip=true]{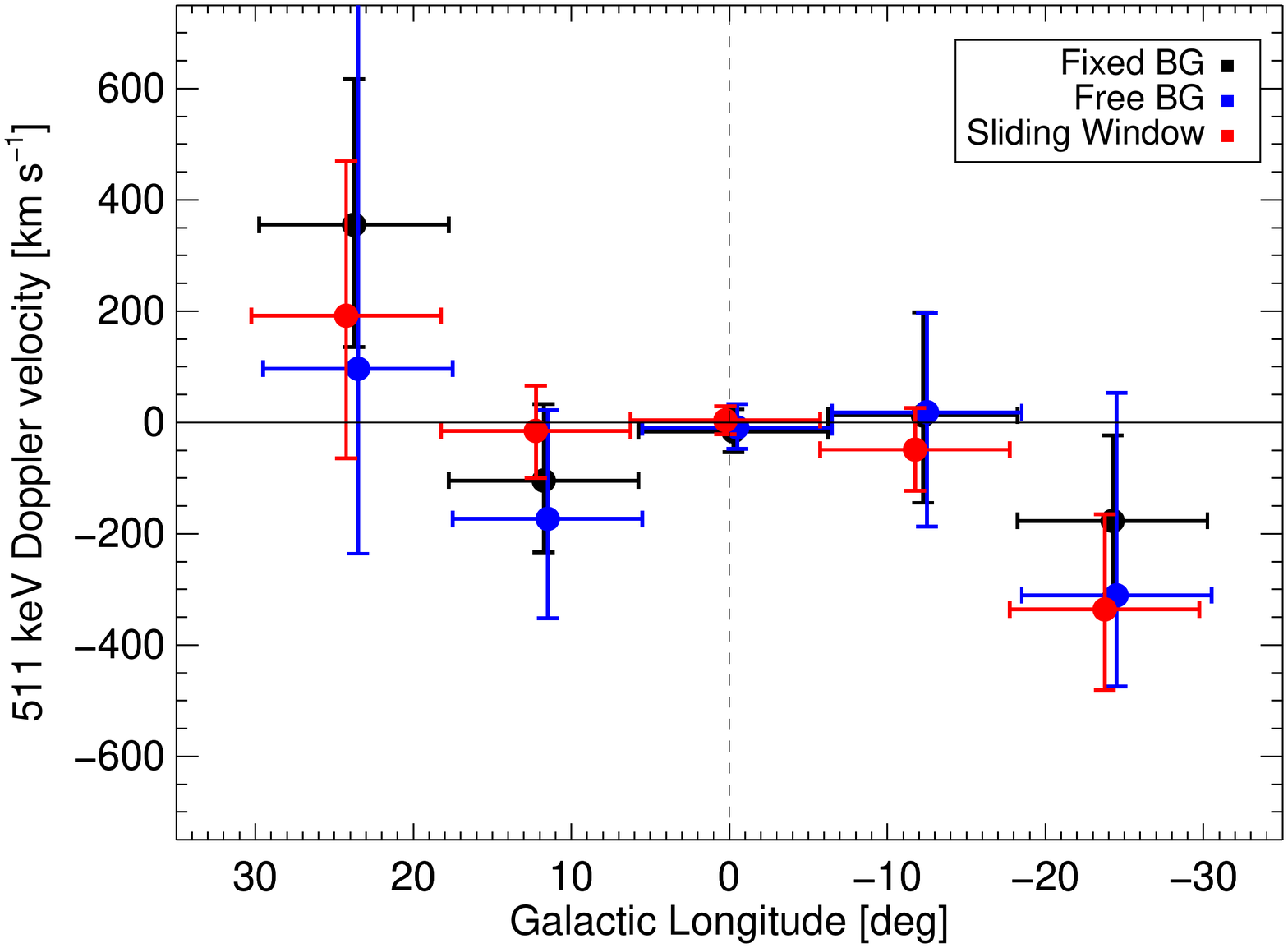}}\\
		  \subfloat[Astrophysical FWHM $\Gamma_L$. \label{fig:compare_fwhm_plot}]{\includegraphics[width=0.89\columnwidth,trim=0.68in 0.72in 0.97in 0.96in,clip=true]{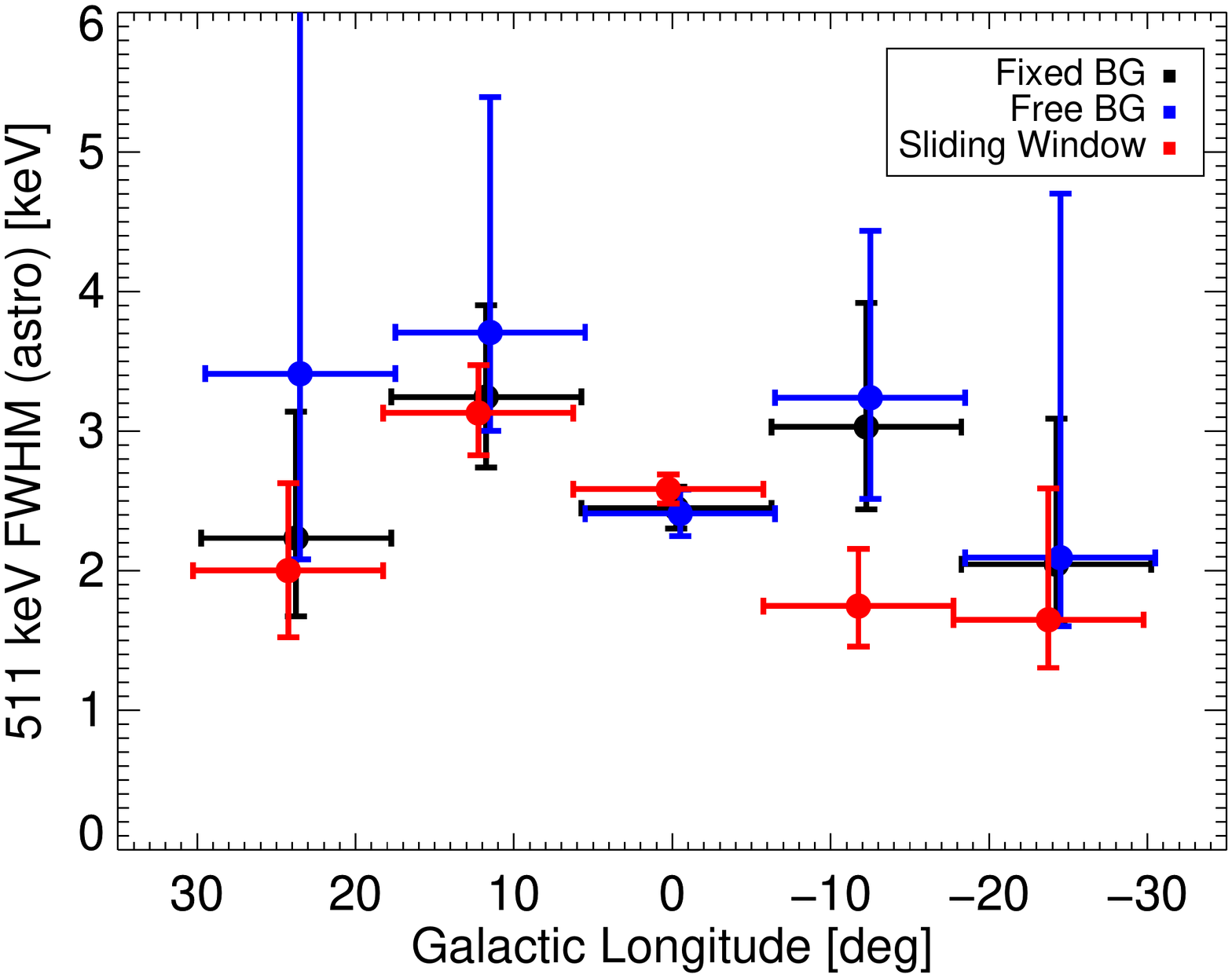}}~
			\subfloat[Positronium fraction $f_{Ps}$. \label{fig:compare_fps_plot}]{\includegraphics[width=0.89\columnwidth,trim=0.68in 0.72in 0.97in 0.96in,clip=true]{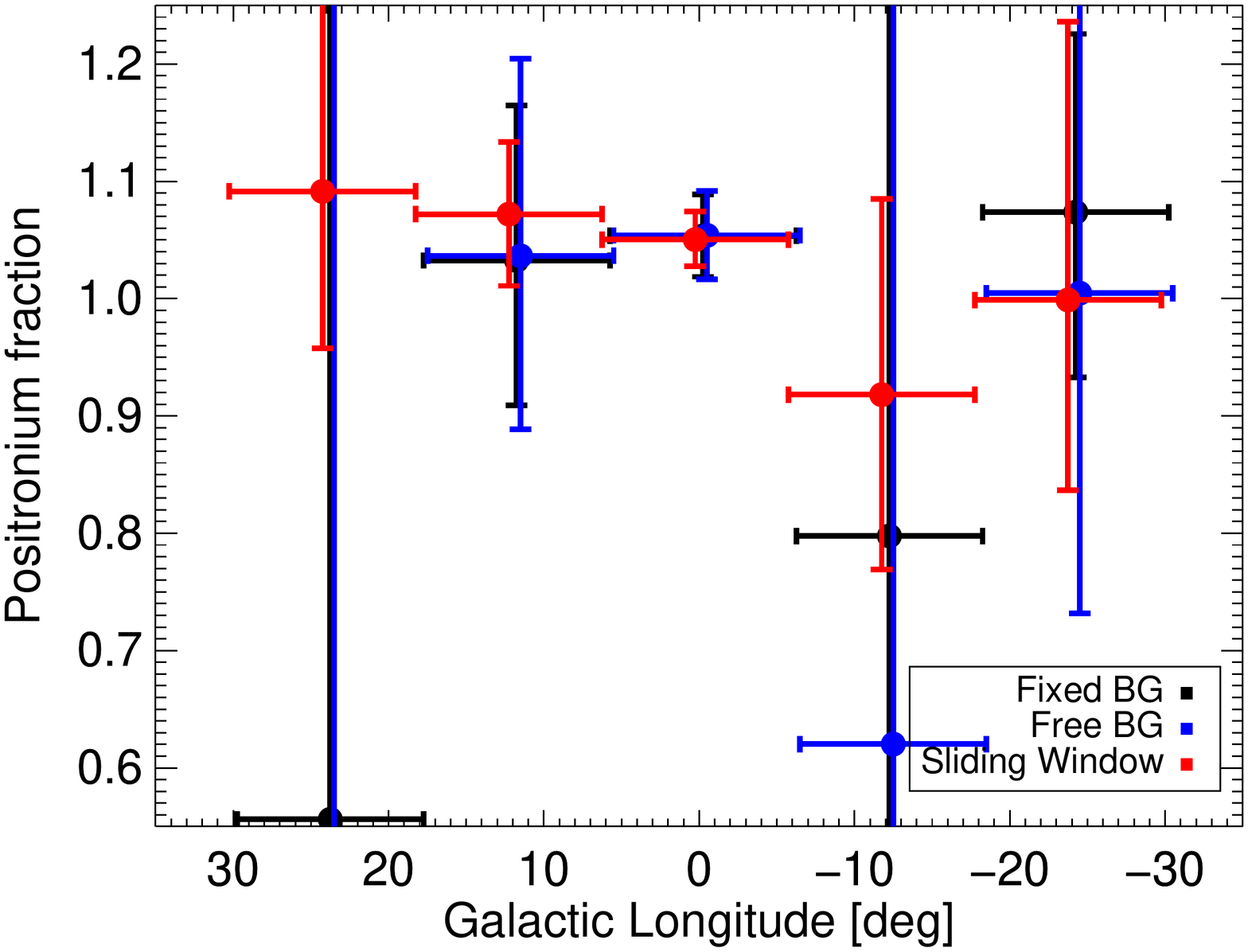}}
	  \cprotect\caption{Derived and fitted spectral parameters ($1\sigma$ uncertainties) from three different analysis methods. The individual data points for each method are shifted in longitude for illustration purpose. The absolute flux values are almost identical. The annihilation parameters, $v_{los}$, $\Gamma_L$, and $f_{Ps}$ are consistent within uncertainties, and show the same trends.}
\end{figure*}

Similar to Figs.~\ref{fig:l_flux_plot}-\ref{fig:l_fps_plot}, we show the spectral parameters derived from the spectra above in Figs.~\ref{fig:compare_flux_plot}-\ref{fig:compare_fps_plot}. A slight difference is found in the Doppler-broadening, as the lines widths in the simultaneous fits appear broader. In general, the parameters coincide and show the same trends along galactic longitudes.

\section{Discussion of systematic uncertainties}\label{sec:systematics}

The follolwing systematics estimates are based on the "sliding window method", Appendix~\ref{sec:slidingwindow}, which obtains up to a factor of two smaller statistical uncertainties. The systematics in the base-line method, therefore, might be larger by this factor.

\begin{figure*}[!ht]
	\centering
		  \subfloat[$+15^{\circ} \geq l \geq +5^{\circ}$. \label{fig:latsys3}]{\includegraphics[width=0.32\textwidth,trim=0.38in 0.49in 0.72in 1.23in,clip=true]{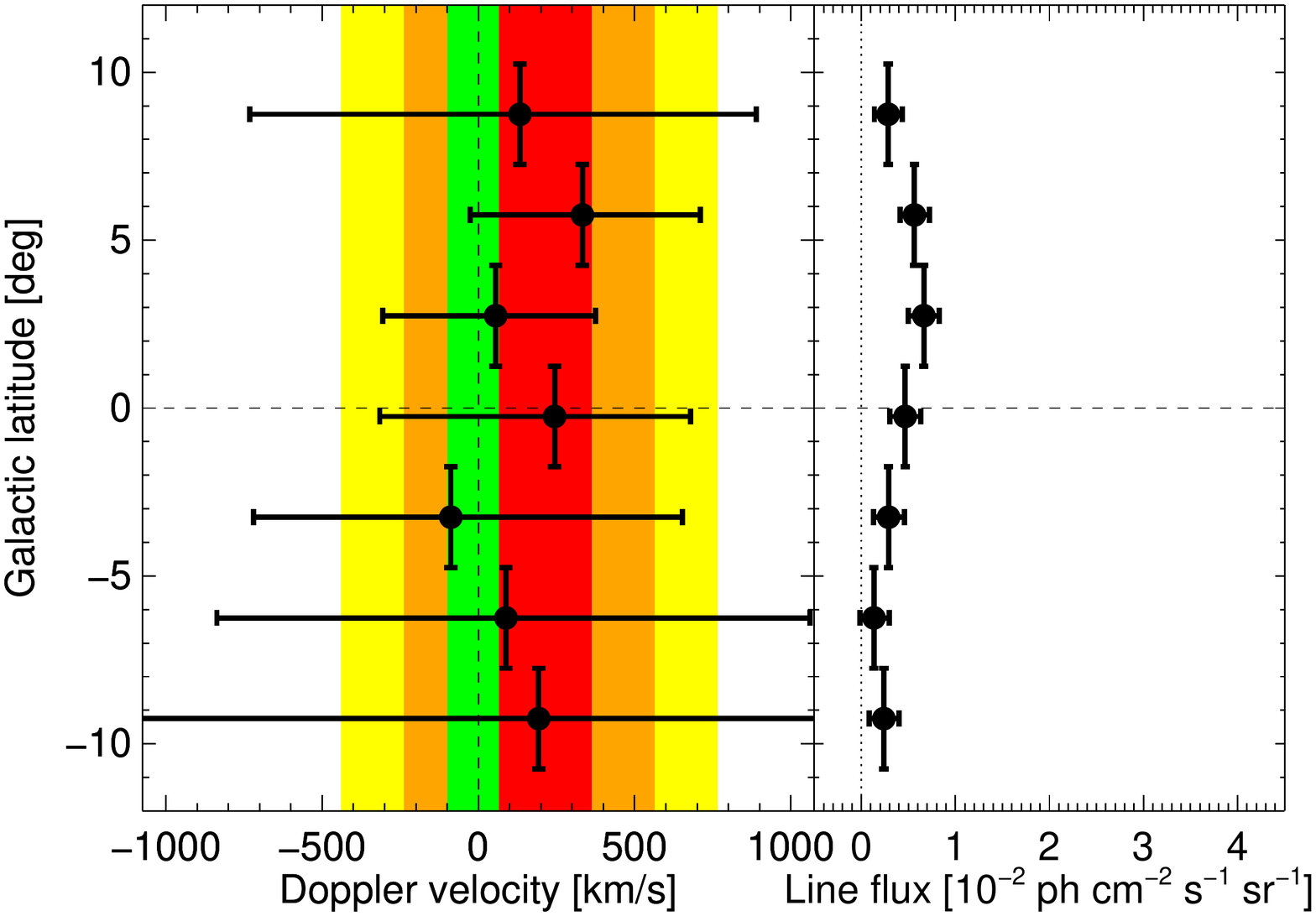}}~
			\subfloat[$+5^{\circ} \geq l \geq -5^{\circ}$. \label{fig:latsys2}]{\includegraphics[width=0.32\textwidth,trim=0.38in 0.49in 0.72in 1.23in,clip=true]{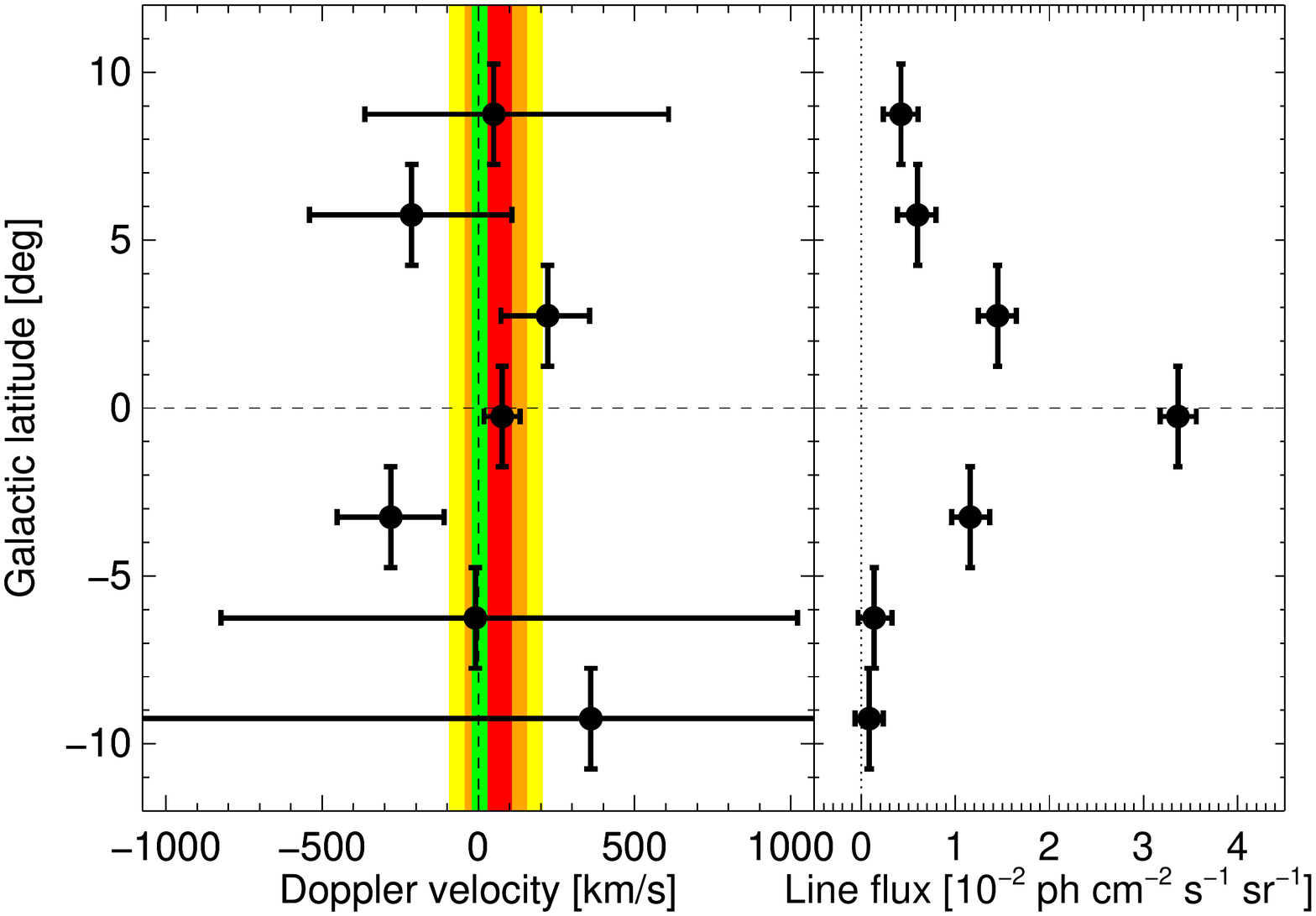}}~
			\subfloat[$-5^{\circ} \geq l \geq -15^{\circ}$. \label{fig:latsys1}]{\includegraphics[width=0.32\textwidth,trim=0.38in 0.49in 0.72in 1.23in,clip=true]{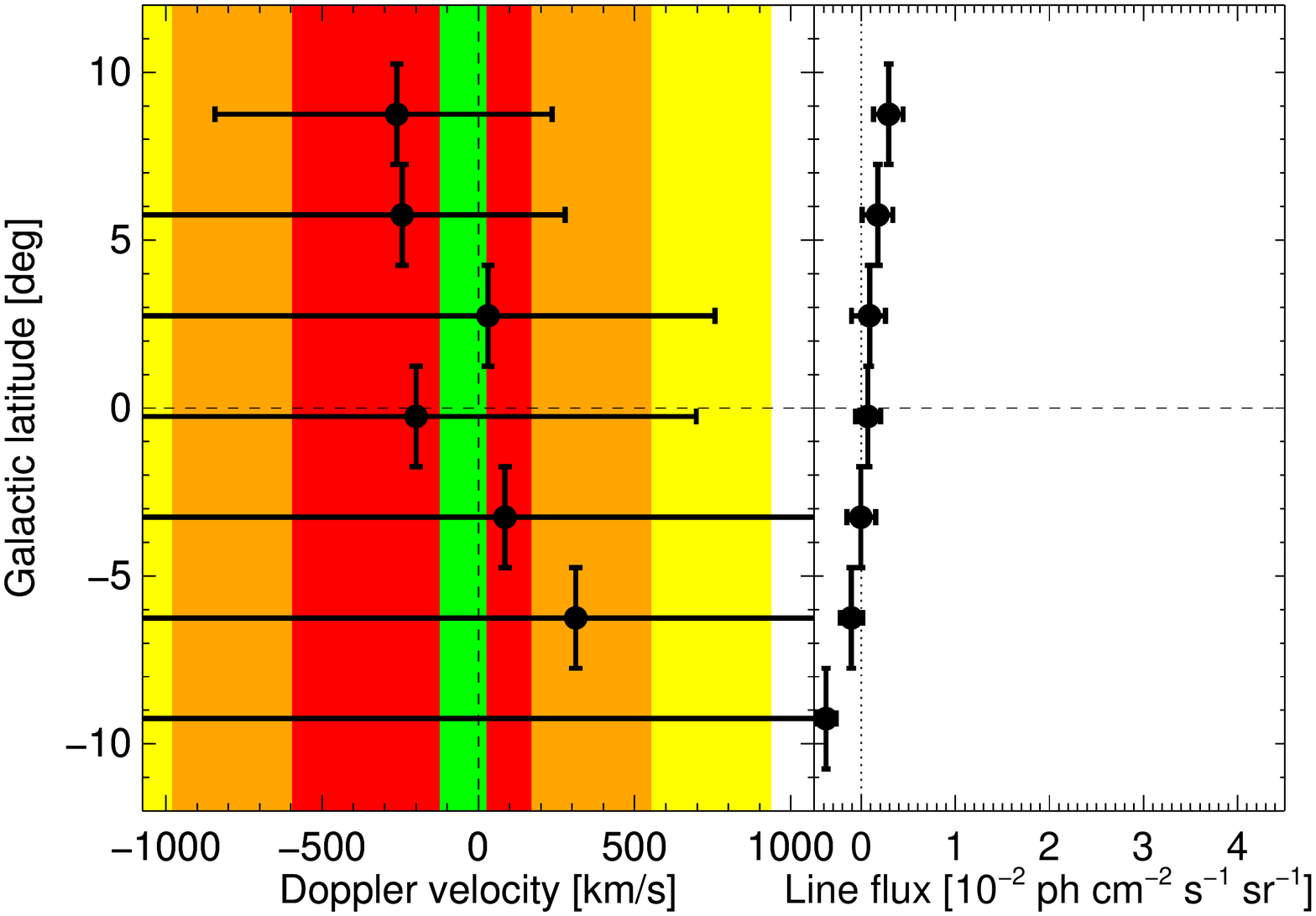}}
	  \caption{Systematic variations in velocity and flux along galactic latitudes and longitudes from the sliding window method. The data points shown are not independent and have partial overlaps in the SPI data space. The left panels of each plot shows the estimated Doppler-velocity as a function of latitude (black data points), together the weighted mean and its 1, 2, and $3\sigma$ uncertainty band in red, orange, and yellow, respectively. The value from the high signal-to-noise spectra, Fig.~\ref{fig:specs_L1}, are marked by a green band. The right panels shows the derived flux of the 511~keV line in each spectrum. See text for more detail.}
\end{figure*}

\citet{Kretschmer2011_PhD} and \citet{Kretschmer2013_26Al} investigated the systematic uncertainties for the l-v-diagram of $\mrm{^{26}Al}$ at 1808.63~keV using different input maps for the maximum likelihood fitting procedure, different ROI sizes, and also different spectral responses to determine the Doppler-shifts. In these previous works, it turned out that for robust estimations of $\gamma$-ray line widths, measured with SPI, and especially at higher energies, a thorough tracing of line degradations with time is required. As a particular sky region may be observed more than once, but at different detector conditions, the measured line width is a superposition of time-weighted detector degradation, and must be considered when estimating the astrophysical line broadening. For Doppler-shifts, on the other hand, the correct response function is not crucial, and a time-variable determination of the bulk velocities lead to a $\lesssim 10~\mathrm{km~s^{-1}}$ systematic effect in the $\mrm{^{26}Al}$ case. Linear extrapolation towards 511~keV would increase this systematic to $\approx 35~\mathrm{km~s^{-1}}$, and is small compared to the statistical uncertainties for longitudes $|l| \gtrsim 6^{\circ}$.

We estimate the effect of different ROI sizes in latitude by performing the same analysis with $\Delta b = 10.5^{\circ}$ and $\Delta b = 21.0^{\circ}$ extents, respectively. Using \verb|S1|, \verb|S2|, and \verb|S3| independently, we find that on average, the velocities are systematically different by about $\approx 25$--$45~\mathrm{km~s^{-1}}$. We also note that the statistical uncertainties are smaller by about $30~\mathrm{km~s^{-1}}$ when using the $\Delta b = 21.0^{\circ}$ ROIs. This indicates weak fluxes at higher latitudes which would support the thick-disk morphology as found in \citet{Siegert2016_511}. Setting a significance cut of $2\sigma$ for the small-ROI sets provides line measurements in the range $-8^{\circ} \lesssim l \lesssim +18^{\circ}$. This asymmetry might arise form the increased line width at positive longitudes, and could be explained by a latidudinal flux variation. Note, that this asymmetry is not the asymmetric longitude-profile as discussed by \citet{Weidenspointner2008_511b}, as our excess appears at positive longitudes. A possible explanation of this asymmetry may be line-of-sight effects of a rotated tri-axial or ellpsoidal bar in the centre of the Milky Way \citep[e.g.][]{Martinez-Valpuesta2011_boxybulgebar,Wegg2013_bulgebar}.

To further test possible latitude variations in flux and Doppler-velocity, we use the 511~keV map and define alternative ROIs to scan the galactic latitude in the same way as described in Appendix~\ref{sec:slidingwindow}. This will also provide an independent check for additional systematic trends in our analysis. These ROIs are three longitude regions, centred at $-10^{\circ}$, $0^{\circ}$, and $+10^{\circ}$, respectively, each with a width of $10^{\circ}$. Between latitudes $-10.75^{\circ}$ and $+10.25^{\circ}$, we construct seven ROIs for each longitude region, for a total of 21 independent ROIs, measuring $\Omega_{sys} = \Delta l_{sys} \times \Delta b_{sys} = 10^{\circ} \times 3^{\circ} \frac{\pi^2}{32400}~\mrm{\frac{sr}{deg^2}} = 0.91\times10^{-2}~\mrm{sr}$. In Figs.~\ref{fig:latsys3} to \ref{fig:latsys1}, the latitudinal variations for the 511~keV Doppler-shifts in the inner Galaxy are shown\footnote{Note that \citet{Siegert2017_PhD} performed this type of latitudinal analysis for the $\mrm{^{26}Al}$ 1.8~MeV line throughout large parts of the Milky Way, and found that increased line-of-sight velocities are possible at higher latitudes from asymmetric superbubble shapes.}.

As the signal-to-noise ratio in these ROIs is rather low, we constrain the spectral fits similar to the sets \verb|S1| to \verb|S3|: the ratio $R_{OL}$ and the width $\sigma$ are fixed to the values obtained from the high signal-to-noise spectra. Only the line fluxes and centroids are determined in the fits. In each latitude subdivision of the ROIs, the derived line-of-sight velocities are consistent with each other (red band), and with the mean of that region (green band). This means that our velocity measurements can be considered robust. The mean flux values as a function of latitude also coincide with the ones derived from the latitude-integrated spectra within $1\sigma$ for longitudes $l \gtrsim -5^{\circ}$, and within $3\sigma$ for $-15^{\circ} < l <-5^{\circ}$. It appears that the flux is distributed asymmetric in latitude over the entire longitude range. This effect could arise, for example, from a nearby annihilation site, such as Scorpius Centaurus \citep[which is detected in $\mrm{^{26}Al}$ and thus a possible site for annihilating positrons;][]{Diehl2010_ScoCen}, or the chosen empirical model to describe the 511~keV emission is not sufficiently-well describing the latidutinal variations in the bulge/bar region. The different exposure time for positive latitudes in these longitude regions cannot account for this effect. Such a positive latitude enhancement is reminiscent of the OSSE map \citep{Purcell1997_511}, showing a fountain-like structure. However, the bin-by-bin analysis performed here is not as extended as the OSSE fountain, but might rather resemble the tilted ellipsoidal shapes, as already observed by \citet{Knoedlseder2005_511}, and reinforces the structure found in that imaging analysis. Using the scatter of the data points in Figs.~\ref{fig:latsys3} to \ref{fig:latsys1}, a systematic uncertainty between $30$ and $150~\mrm{km~s^{-1}}$, depending on the longitude range, is derived.

An additional source of systematic uncertainties may be an unstable calibration of the Ge detector spectra. In Fig.~\ref{fig:calibbgpeaks511}, we show the calibration accuracy of the 511~keV background line as measured with SPI over the ongoing INTEGRAL mission, up to orbit number 1593 (2015, Oct 1). Clearly, the calibration is accurate in the region around 511~keV to about 0.05~keV, corresponding to a two-sided systematic uncertainty in the velocity measurement of $\pm 15~\mrm{km~s^{-1}}$. Another systematic offset of 0.035~keV towards lower energies can be identified, shifting the absolute velocity by about $20~\mrm{km~s^{-1}}$. A weak long-term variation can also be seen, but which can only account for $3~\mrm{km~s^{-1}}$. In total, the calibration may account for up to $\approx 40~\mrm{km~s^{-1}}$ of systematics. From the partially overlapping ROI sets \verb|L1| to \verb|L4|, we illustrate the systematic uncertainties arising for all spectral fit and derived parameters in Appendix~\ref{sec:systematic_trend}. In fact, considering the non-independent data points can reveal an underlying structure, but which has to be taken with caution. From the scatter of data points covering the same regions, we estimate additional systematic uncertainties:

The total systematic uncertainty depends on the longitude range, and we estimate of the order of $100~\mrm{km~s^{-1}}$ for $|l|\lesssim^{\circ}$, and $200~\mrm{km~s^{-1}}$ for $|l|\gtrsim5^{\circ}$, respectively, including all above-described effects. For the Ps fraction, $f_{Ps}$, we find $\lesssim 0.05$ systematics for all longitudes. The FWHM varies $\approx 0.05$ to $\approx 0.3$~keV for small longitudes, and up to 0.6~keV near $|l|\approx 30^{\circ}$. Systematic deviations of the o-Ps flux are between $0.2$ and $0.5 \times 10^{-3}~\mrm{ph~cm^{-2}~s^{-1}}$. Similarly, the line flux variations are below $\lesssim 0.5 \times 10^{-4}~\mrm{ph~cm^{-2}~s^{-1}}$. The systematic uncertainties on the spectral parameters like centroid and width are smaller in the inner Galaxy than in the disk, which is expected from the higher accuracy in largely exposed regions. However, the apparent systematics in the integrated fluxes are larger for the inner Galaxy than in the disk (see Appendix~\ref{sec:systematic_trend}), which would only be expected if an underlying astrophysical variation was present. We therefore conclude, that the systematic flux variations as a function of longitude originate from gradients on smaller scales as the bin size of $\Delta l = 12^{\circ}$. As a result, this means that the o-Ps as well as the 511~keV line flux show larger fluxes at positive longitudes then what is seen at negative longitudes. The flux analysis in the small ROI sets independently validates this conclusion.

\section{Spectral trends along galactic longitudes}\label{sec:systematic_trend}

\begin{figure*}[!ht]
	\centering
		  \subfloat[Integrated 511~keV line flux $F_L$. \label{fig:l_flux_sys}]{\includegraphics[width=0.32\textwidth,trim=0.82in 0.64in 0.91in 0.64in,clip=true]{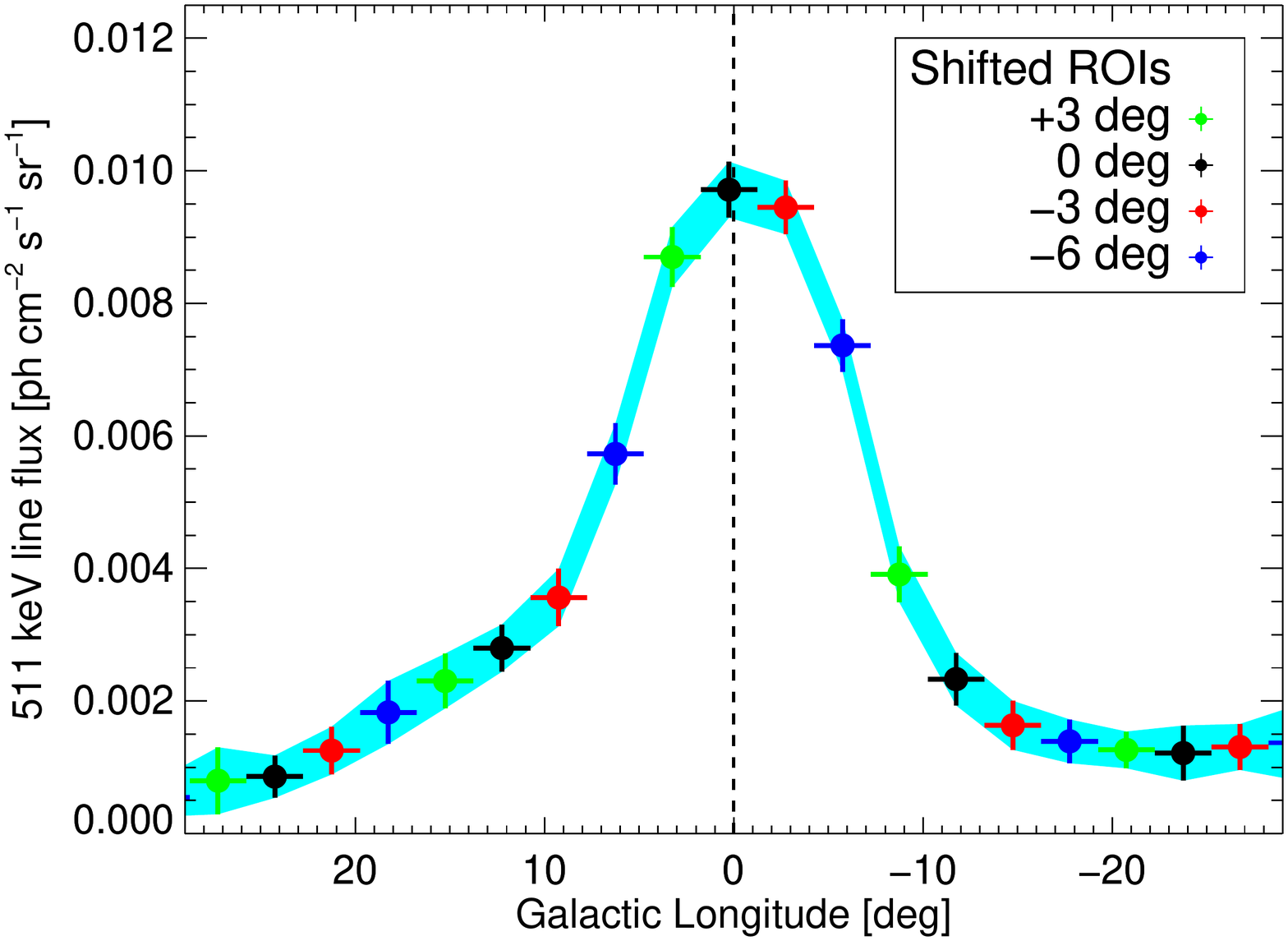}}~
			\subfloat[Line of sight Doppler-velocity $v_{los}$. \label{fig:l_v_sys}]{\includegraphics[width=0.32\textwidth,trim=0.82in 0.64in 0.91in 0.64in,clip=true]{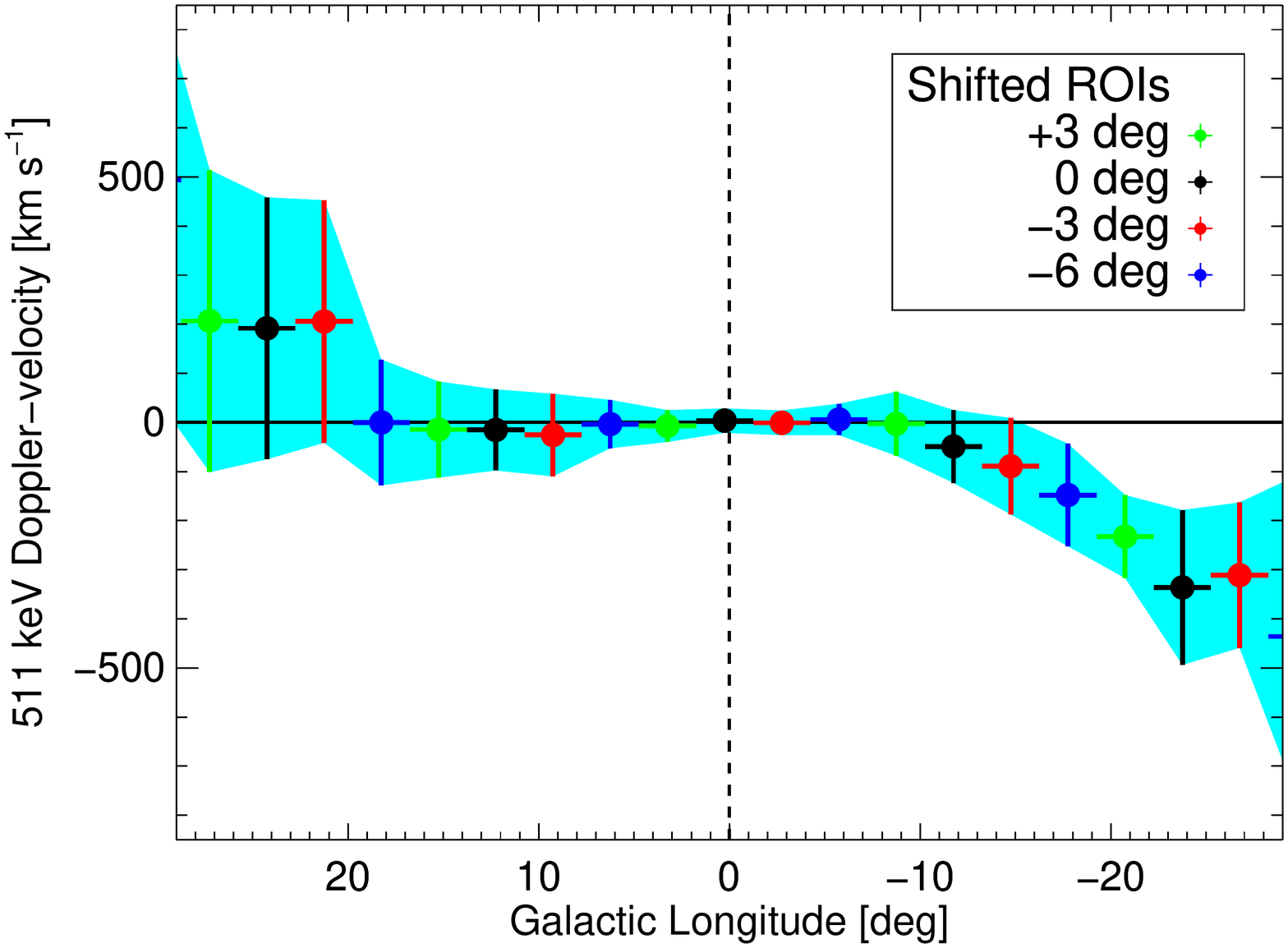}}~			
			\subfloat[o-Ps flux $F_O$. \label{fig:l_ops_sys}]{\includegraphics[width=0.32\textwidth,trim=0.82in 0.64in 0.91in 0.64in,clip=true]{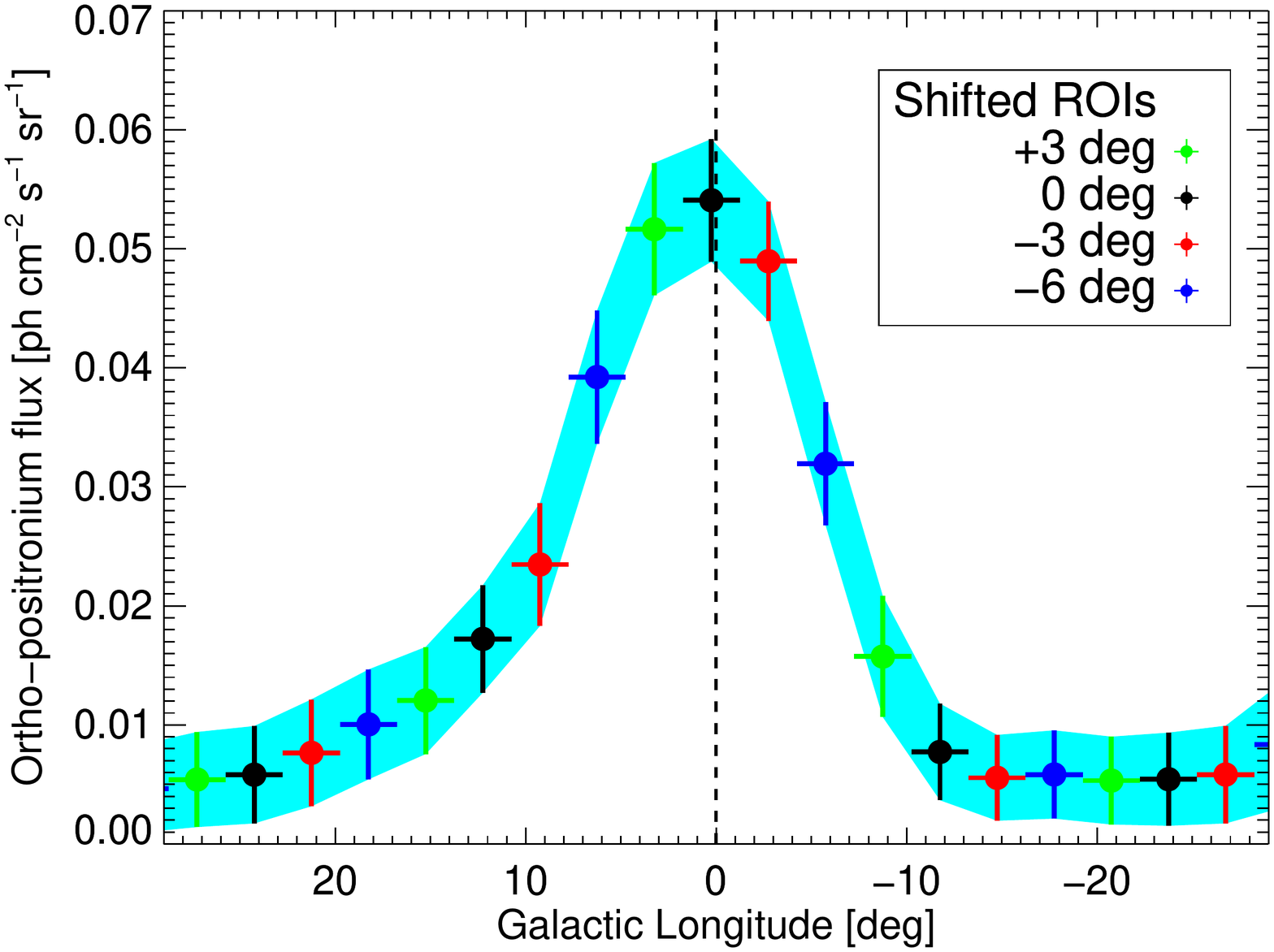}}\\
			\subfloat[Positronium fraction $f_{Ps}$. \label{fig:l_fps_sys}]{\includegraphics[width=0.32\textwidth,trim=0.82in 0.64in 0.91in 0.64in,clip=true]{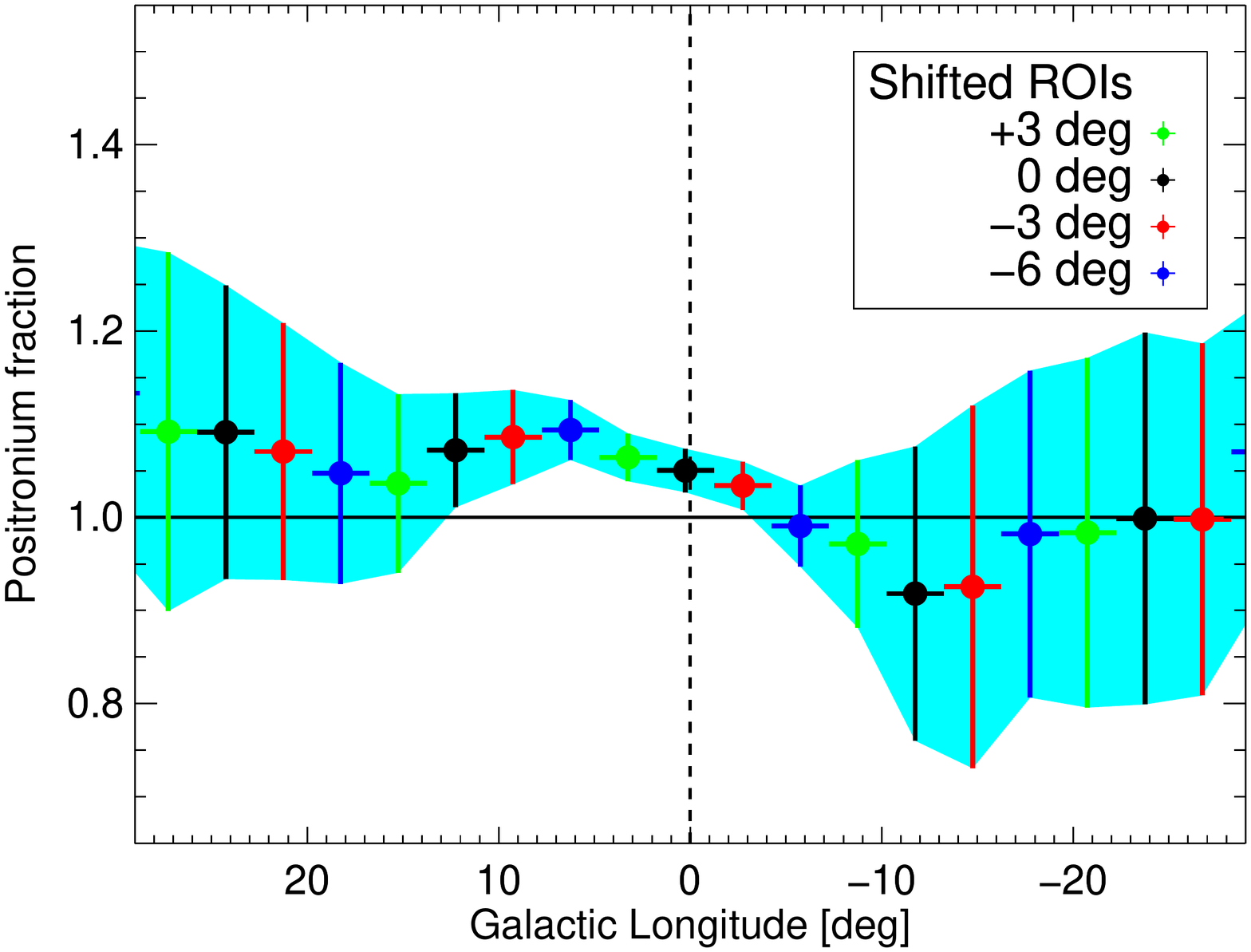}}~
		  \subfloat[Astrophysical FWHM $\Gamma_L$. \label{fig:l_fwhm_sys}]{\includegraphics[width=0.32\textwidth,trim=0.82in 0.64in 0.91in 0.64in,clip=true]{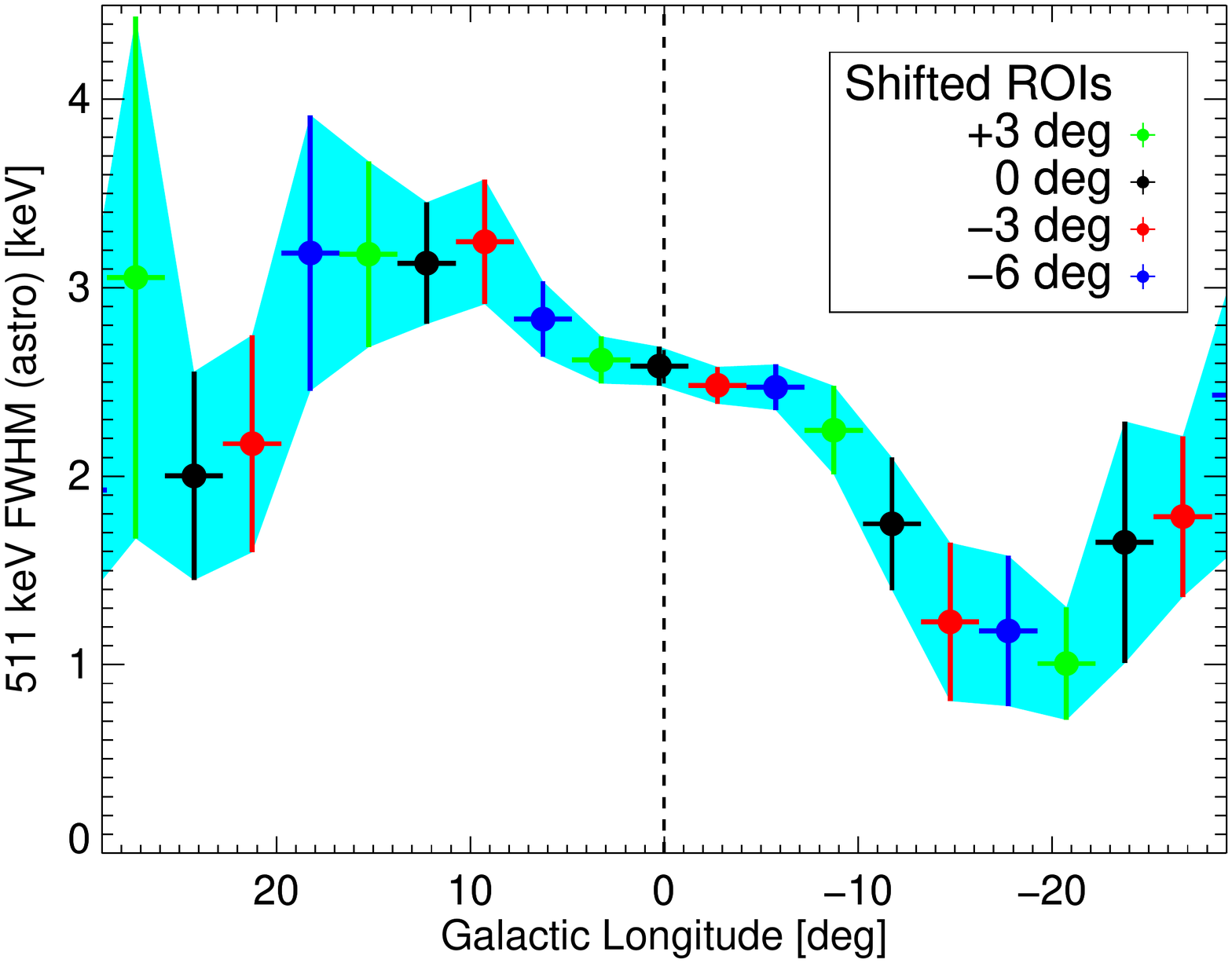}}~
			\subfloat[Continuum flux density $C_0$. \label{fig:l_conti_sys}]{\includegraphics[width=0.32\textwidth,trim=0.82in 0.64in 0.91in 0.64in,clip=true]{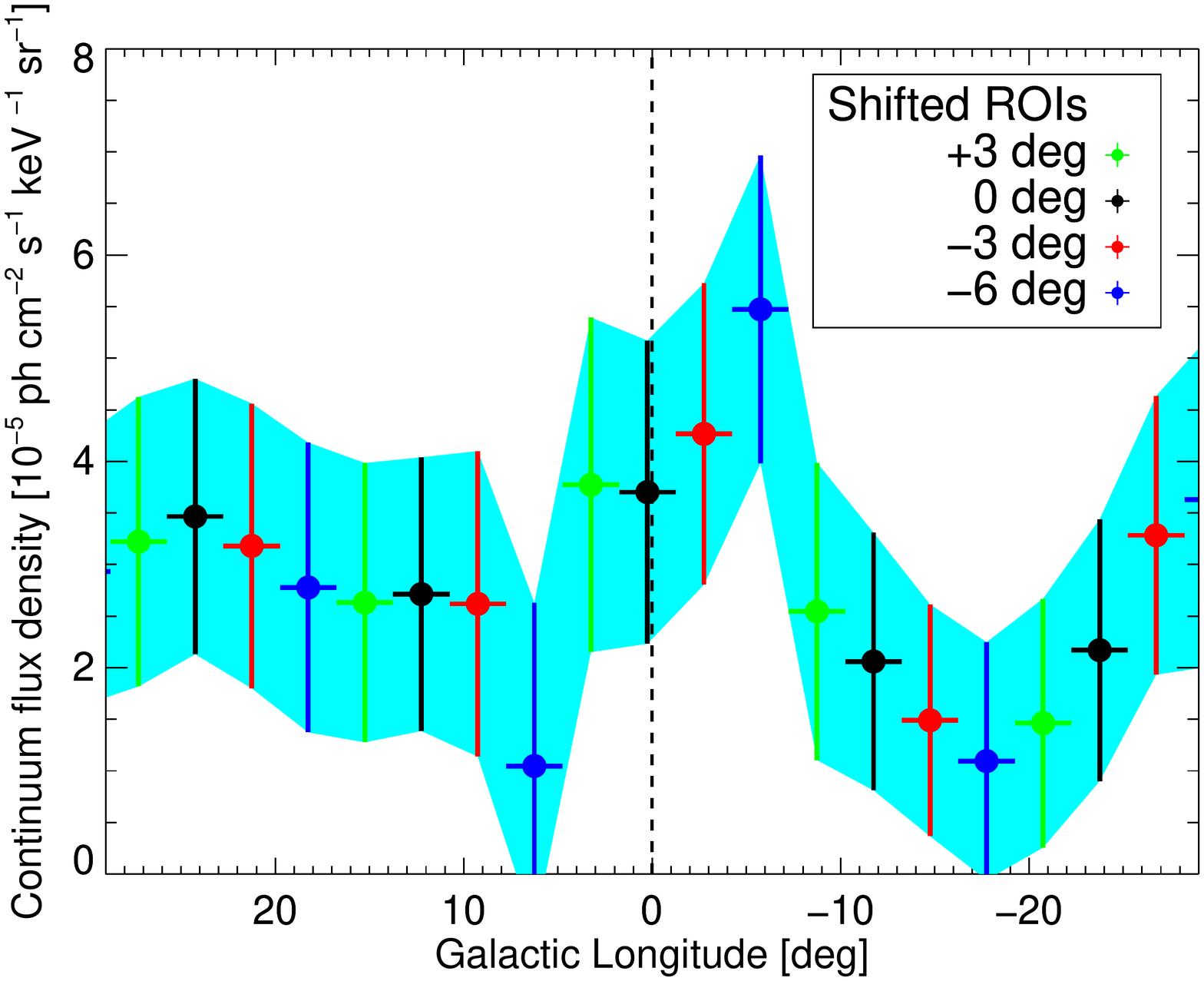}}
	  \cprotect\caption{Systematics trends along galactic longitudes from overlapping ROIs. The resulting data points are not independent and have partial (same colours) to substantial (alternating colours) overlaps in the SPI data space. See text for details.}
\end{figure*}

The choice of the ROI set to extract spectral parameters is not unique. Therefore, we performed spectral fits of partially overlapping regions to estimate systematic deviations from a chosen set using the "sliding window method". This might in addition also reveal underlying trends as a function of longitude. In Figs.~\ref{fig:l_flux_sys} to \ref{fig:l_conti_sys}, the fitted and derived spectral parameters for the overlapping ROI sets \verb|L1| to \verb|L4| are shown. Each data point in the figures covers a solid angle of $\Omega_L = \Delta l \times \Delta b = 12^{\circ} \times 21^{\circ}$. For illustration purpose, we show smaller error bars on the x-axes, and draw an uncertainty band, connecting each upper and lower error bar. By combining four data points which partially cover the same regions, we can estimate the systematic uncertainties as a function of longitude, which results from the observation time of this region in the sky and the actual signal strength.

We distinguish between two prominent regions to evaluate the systematic uncertainties, the inner bar or bulge region at $-18^{\circ} \lesssim l \lesssim +18^{\circ}$, and the disk outside this area. The Doppler-velocity shows systematic trends below $\lesssim 20~\mrm{km~s^{-1}}$ in the bulge, and rises up to $\approx 120~\mrm{km~s^{-1}}$ in the disk. While the FWHM varies systematically between 0.05 and 0.3~keV in the bulge, the variation may be as large as 0.6~keV in the disk. The Ps fraction shows variations everywhere below 0.05. On average, the continuum flux density shows systematics of the order of $10^{-6}~\mrm{ph~cm^{-2}~s^{-1}~keV^{-1}}$ in the inner Galaxy, and below $5 \times 10^{-7}~\mrm{ph~cm^{-2}~s^{-1}~keV^{-1}}$ in the disk. The o-Ps flux can vary between $0.5$ and $1.5 \times 10^{-3}~\mrm{ph~cm^{-2}~s^{-1}}$ in the bulge region, and the systematic deviations are below $0.2 \times 10^{-3}~\mrm{ph~cm^{-2}~s^{-1}}$ outside the bulge. Likewise, the line flux variations in the bulge region are larger with $0.5$--$2.5 \times 10^{-4}~\mrm{ph~cm^{-2}~s^{-1}}$ than in the disk with below $0.5 \times 10^{-4}~\mrm{ph~cm^{-2}~s^{-1}}$.

It is evident that the derived spectral parameters, such as centroid and width, show small variations in the centre and large variations at larger longitudes. This is expected from the decreasing observation time and signal strength for increasing $|l|$. However, the flux amplitudes show the opposite behaviour, with apparently large variations in the bulge and small in the disk. As this is unexpected from the exposure time considerations, we conclude that the flux variations in the bulge region are not systematic effects from the choice of the binning but reflect more the true morphological profiles in each spectral feature. This means that the line as well as the o-Ps flux show an enhancement for positive longitudes with respect to the same position at negative longitudes. As discussed in the main text, this would be expected from an ellipsoidal and tilted bar as observed from infrared light in the Milky Way.

\section{Details on the spatial emission model for positron annihilation}\label{sec:appendix_morphology}

In total, the $\gamma$-ray emission in the range 490--530~keV is modelled by the sum of the four 2D-Gaussian components, as

\begin{equation}
M_{511}(l,b) = A_{tot} \times \sum_{i=1}^{4} \frac{F^i}{2\pi\sigma_l^i\sigma_b^i} \exp \left( - \frac{(l - l_C^i)^2}{2\sigma_l^{i2}} - \frac{(b - b_C^i)^2}{2\sigma_b^{i2}} \right)\mrm{,}
\label{eq:gauss_2d2}
\end{equation}

with the parameters for each component $i=\{\mrm{NB,BB,GCS,DISK}\}$ provided in Tab.~\ref{tab:model_components}. In Eq.~(\ref{eq:gauss_2d2}), $F^i$ is the relative flux (relative weighting) of each 2D-Gaussian component, centred at $(l_C^i/b_C^i)$, with widths $(\sigma_l^i,\sigma_b^i)$ in longitude and latitude, respectively. The absolute amplitude $A_{tot}$ is determined in the maximum likelihood fit, and corresponds to a value $\theta_i$ in Eq.~(\ref{eq:modeldesc}).

\begin{table}[!ht]
\centering                                      
\begin{tabular}{l rrrrr}          
\hline                        
Comp. & $F$ & $l_C$ [deg] & $b_C$ [deg] & $\sigma_l^{\Gamma}$ [deg] & $\sigma_b^{\Gamma}$ [deg] \\
\hline                                   
    NB   &  $0.32$                     & $-1.25$   & $-0.25$ & $5.75$   & $5.75$  \\      
    BB   &  $0.64$                     & $0.00$    & $0.00$  & $20.55$  & $20.55$ \\
		GCS  &  $0.08$                     & $-0.06$   & $-0.05$ & $0.00$   & $0.00$  \\
    DISK &  $1.66$                     & $0.00$    & $0.00$  & $141.29$ & $24.73$ \\
\hline                                             
\end{tabular}
\caption{Characteristics of the sky model components in the maximum likelihood model fitting analysis \citep[adapted from][]{Siegert2016_511}. The values $\sigma_{(l/b)}^{\Gamma}$ are the FWHM of Gaussian components, calculated via $\sigma_{(l/b)}^{\Gamma} = 2\sqrt{2\ln(2)}\sigma_{(l/b)}$.}              
\label{tab:model_components}      
\end{table}


\section{Spectral fit adequacy}\label{sec:corner_plots}

\begin{figure*}[!ht]
	\centering
		  \subfloat[$-5.75^{\circ} \geq l \geq +6.25^{\circ}$. \label{fig:corner_plot03}]{\includegraphics[width=0.85\columnwidth,trim=0.09in 0.85in 0.30in 0.88in,clip=true]{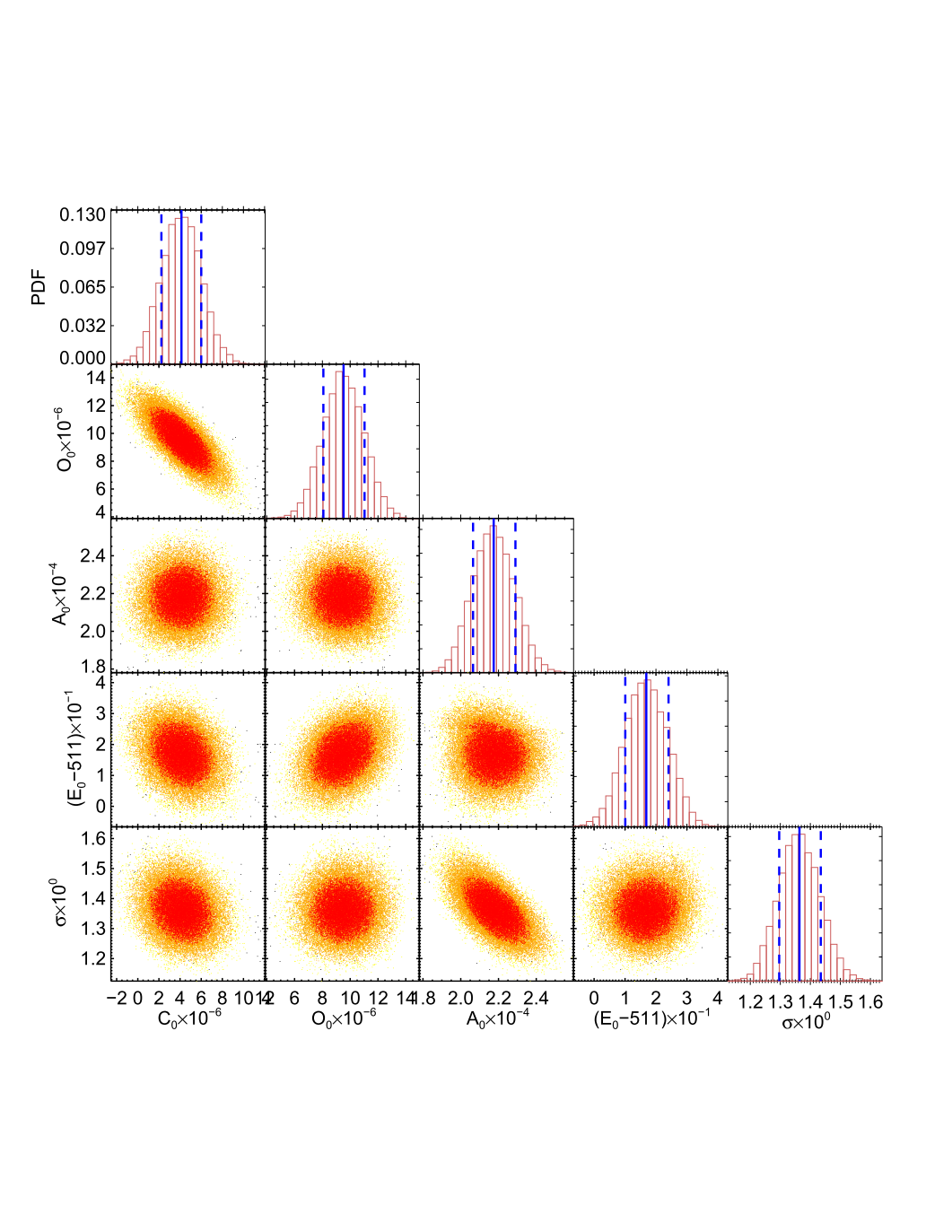}}~
			\subfloat[$-29.75^{\circ} \geq l \geq -17.75^{\circ}$. \label{fig:corner_plot05}]{\includegraphics[width=0.85\columnwidth,trim=0.09in 0.85in 0.30in 0.88in,clip=true]{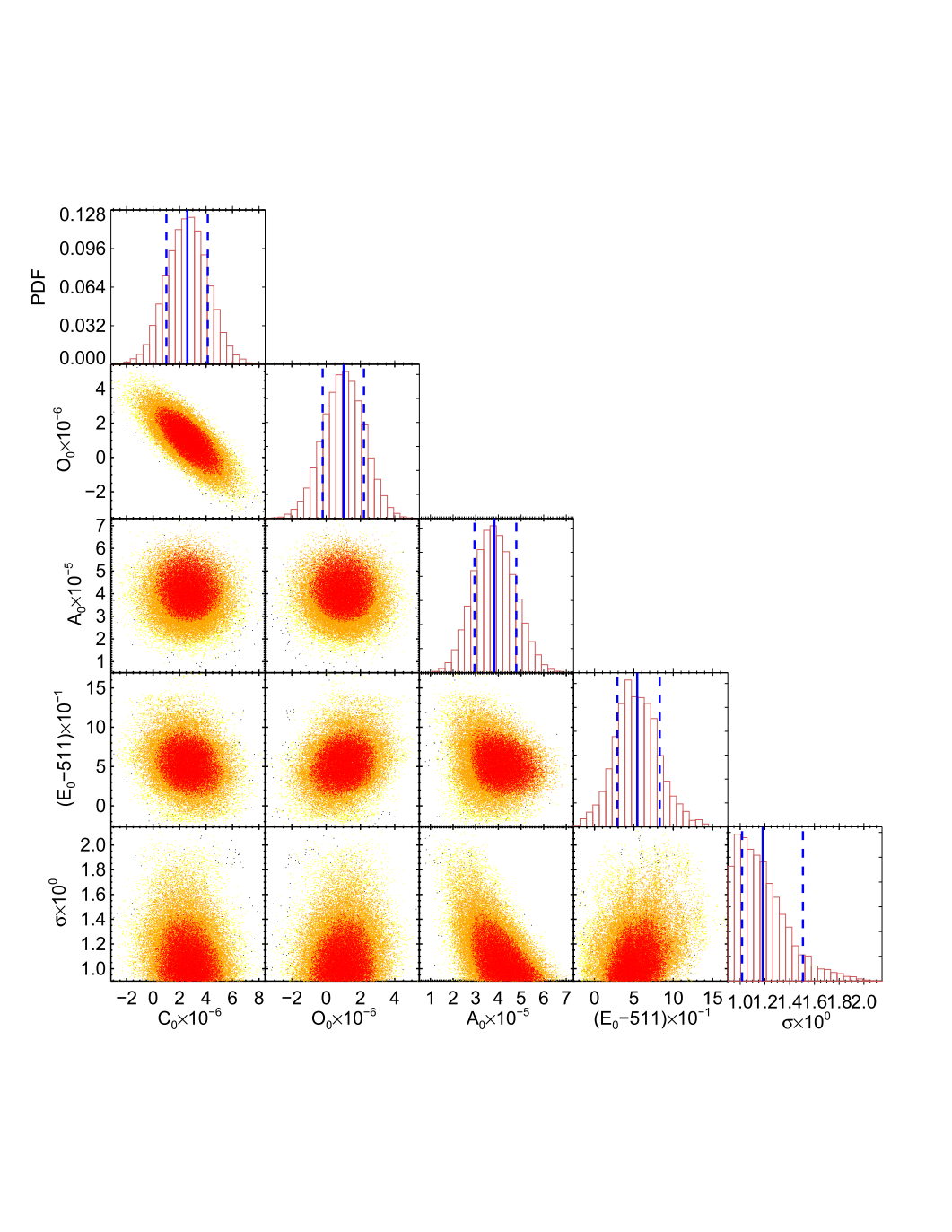}}
	  \caption{Corner plots of spectral fit parameters in chosen ROIs. In each 2D plot, the joint posterior distributions are shown, colour-coded by the likelihood, with red, orange, and yellow showing the 1, 2, and 3$\sigma$ regions. On the diagonals, the marginalised PDFs of each parameter are shown. The solid blue line marks the expectation (=fit) value for the parameter, and the dashed blue lines the 68.3\% intervals. Note the natural anti-correlations between the continuum and o-Ps amplitude ($C_0$ vs. $O_0$), and between the line amplitude and width ($A_0$ vs. $\sigma$).}
\end{figure*}


In Figs.~\ref{fig:corner_plot03} and \ref{fig:corner_plot05}, the corner plots of the Monte Carlo Markov Chain fits of the central (high signal-to-noise) and the most blue-shifted (low signal-to-noise) spectrum, respectively, are shown. In each spectrum, five parameters are fitted, with each parameter having a flat prior distribution. The continuum and line amplitude $C_0$ as well as the line amplitude $A_0$ may range between $-\infty$ to $+\infty$, the o-Ps continuum amplitude $O_0$ is constrained to positive values, and the width $\sigma$ may not be smaller than 0.91~keV (instrumental resolution). In each fit, the centroids of the symmetric Gaussian $E_0$ is constrained between 509 and 513~keV, corresponding to Doppler-velocities of $\pm 1200~\mrm{km~s^{-1}}$. The fit parameters and uncertainties have been determined from the marginalised probability density functions (PDFs), calculating the expectation value as the fit value, and the 68.3\% intervals around that value for the uncertainties.

\section{Line shape variations from nucleosynthesis positrons}\label{sec:subMeV}

\begin{figure}[!ht]%
\centering
\includegraphics[width=0.9\columnwidth,trim=1.4cm 1.8cm 2.3cm 2.1cm,clip=true]{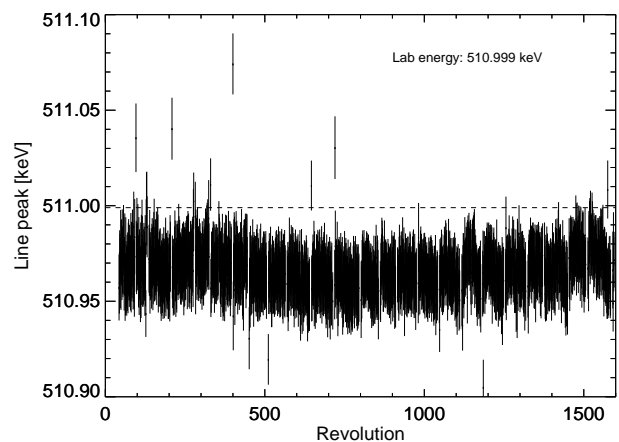}%
\caption{Measured line peak of the SPI background 511~keV line. Each data point (vertical bar, with uncertainty) represents the line peak, determined via Eq.~(\ref{eq:epeak_full}), of all detectors in one orbit (revolution) of the INTEGRAL mission. The data set used ranges up to revolution 1279. Outliers, such as around revolution 400 ($E_{peak} = 511.075\pm0.015~\mrm{keV}$) have been excluded from our data set beforehand. The horizontal dashed line marks the laboratory energy.}%
\label{fig:calibbgpeaks511}%
\end{figure}

Direct annihilation (= annihilation in flight) would reduce the intensity of o-Ps photons in the disk. For example at 1~keV, about 90\% Ps and 10\% inflight-annihilations of positrons with free electrons would be expected. This ratio, however, strongly depends on the available species in an HI cavity (bubble), and may fluctuate between bubbles. The Ps fraction in the disk could well be lower than in the bulge, owing to the large uncertainties. \citet{Siegert2016_511} found $f_{Ps}^{disk} = 0.90\pm0.19$ for the disk as a whole, similar to what we find for the $|l| \gtrsim 6^{\circ}$ bins. Within $3\sigma$, values as small as 0.3--0.4 are thus possible. Thus, a fraction of positrons may annihilate in flight at large velocities. The resulting line-of-sight velocities of positron annihilation inside bubble walls would show a preferred direction: the annihilation luminosity inside the bubble walls would be uniform, but the flux from the distant parts will be reduced due to the increased distance ($r^{-2}$-law). This leads to an apparent blue-shift from all directions, independent of galactic rotation. This shift would add to the average $\mrm{^{26}Al}$ velocity, something that is not seen in the data. The $3\sigma$-limit on the blue-shifted bulk-motion is about $60$--$80~\mrm{km~s^{-1}}$ (see Tab.~\ref{tab:lvchi2}), so that annihilation in flight at non-relativistic velocities probably only makes a small contribution to the measured line profiles. A systematic uncertainty in the calibration of the SPI spectra could also lead to a one-sided over-estimation, i.e. many values would either be on the blue side \emph{or} the red side, not equally distributed. As shown in Fig.~\ref{fig:calibbgpeaks511}, the variations in the energy calibration of SPI are only a small part of our velocity measurements (for a full discussion of systematic uncertainties, see Appendix~\ref{sec:systematics}).

\section{Functional forms of spectral components and other derived parameters}\label{sec:appendix_functions}

The full expression of the functions from Eqs.~(\ref{eq:conti})-(\ref{eq:ortho_function}) are:

\begin{align}
C(E;C_0,\alpha) & = C_0 \left(\frac{E}{511~\mrm{keV}}\right)^{\alpha} \label{eq:conti_full} \\
\cline{1-2}
G(E;A_0,E_0,\sigma) & = A_{0} \exp \left(- \frac{(E-E_{0})^2}{2\sigma} \right) \label{eq:gaussian_full} \\
T(E;\tau) & = \frac{1}{\tau} \exp \left( - \frac{E}{\tau} \right) \quad \forall E > 0 \label{eq:tail_full} \\
L(E;A_0,E_0,\sigma,\tau) & = (G \otimes T)(E) = \nonumber \\
& = \sqrt{\frac{\pi}{2}} \frac{A_{0} \sigma}{\tau} \exp \left( \frac{2 \tau (E-E_{0}) + \sigma^2}{2 \tau^2} \right) \nonumber \\
& \erfc \left( \frac{\tau (E-E_{0}) + \sigma^2}{\sqrt{2} \sigma \tau} \right) \label{eq:cls_function_full} \\
\cline{1-2}
O(E;O_0,E_0) & = 2 O_0 \left( T_1 - T_2 \ln(T_3) + T_4 + T_5 \ln(T_3) \right)\mrm{,~with}\label{eq:ortho_function_full} \\
T_1 & = \frac{E(E-E_0)}{(2E_0-E)^2} \\
T_2 & = \frac{2E_0(E_0-E)^2}{(2E_0-E)^3} \\
T_3 & = \frac{E_0-E}{E_0} \\
T_4 & = \frac{2E_0-E}{E} \\
T_5 & = \frac{2E_0(E_0-E)}{E^2}
\end{align}

The integrated flux $I_L$ of the line function $L(E)$, Eq.~(\ref{eq:cls_function_full}), is given by the integral over that convolved line shape. Thanks to the conservation of area during the convolution, the line flux of the degraded Gaussian $L(E)$ is identical to that of the symmetric Gaussian $G(E)$, namely

\begin{align}
I_L & = \int_{-\infty}^{+\infty} L(E) dE = \int_{-\infty}^{+\infty} (G \otimes T)(E) dE = \nonumber\\
& = \int_{-\infty}^{+\infty} G(E) dE = \sqrt{2\pi} A_0 \sigma\mrm{.} \label{eq:intensity_constant} 
\end{align}

Note that the line amplitude $A_0$ and width $\sigma$ are anti-correlated, which must be taken into account when then uncertainties for $I_L$ are propagated. Similarly, the integrated flux of the o-Ps continuum $I_O$ is given by

\begin{equation}
I_O = \int_{0}^{E_0} O(E) dE = O_0 E_0 (\pi^2 - 9)\mrm{.}
\label{eq:ortho_int}
\end{equation}

The Ps fraction, $f_{Ps}$, is the relative amount of positrons forming the bound Ps state with respect to annihilating directly with electrons. It is calculated from the statistical weight of Ps decaying into two or three photons, and the number of direct annihilations. Ps is either formed in the para (p) state with total spin $S=0$ or the ortho (o) state with $S=1$. The multiplicity of a particular spin state is $(2S+1)$, so that p-Ps will be formed $1/4$ of the time, and o-Ps $3/4$ of the time. While p-Ps decays into two photons similar to direct annihilation, o-Ps will result in three photons. Consequently, the o-Ps flux, $I_O$, will be proportional to $3\times3/4f_{Ps} = 9/4 f_{Ps}$, and the two-photon flux, $I_L$, will be proportional to $2\times1/4f_{Ps} + 2\times(1-f_{Ps}) = 2 - 3/2f_{Ps}$. From a measurement of the total annihilation $\gamma$-ray intensity, the Ps fraction is given by

\begin{equation}
f_{Ps} = \frac{8R_{OL}}{6R_{OL}+9}\mrm{,}
\label{eq:fps_definition}
\end{equation}

where $R_{OL}$ is the ratio between the three-photon flux and the two-photon flux, $R_{OL} = I_O/I_L$. Finally, if all positrons annihilate via Ps formation, $f_{Ps}$ will be 1.0, corresponding to a maximum flux ratio $R_{OL}^{max} = 4.5$.

The FWHM, $\Gamma_L$ of the degraded lines can be approximated by	

\begin{equation}
\Gamma_L \approx \Gamma \left[a_0 + \sqrt{(1 - a_0)^2 + \left(\frac{a_1 \tau}{\Gamma}\right)^2}\right]\mrm{,}
\label{eq:fwhm_line}
\end{equation}

where $\Gamma = 2\sqrt{2\ln2}\sigma$ is the normal FWHM of the symmetric Gaussian $G(E)$, and $a_0 = 0.913735$ and $a_1 = 0.710648$ are constants \citep{Kretschmer2011_PhD}.

\end{document}